\documentclass[preprint]{aastex}
\usepackage{epsfig,emulateapj5}

\newcommand{\km}{${\rm km\,s}\, ^{-1}$}
\newcommand{\fuse}{{\em FUSE}}

\shorttitle{{\em FUSE}\/ Observations of the Magellanic Bridge gas}
\shortauthors{Lehner}
 \begin{document}
\title{{\em FUSE}\/ Observations of the Magellanic Bridge Gas toward Two Early-Type Stars: 
Molecules, Physical Conditions, and Relative Abundances}

\author{N. Lehner}
\affil{The Johns Hopkins University, Department of Physics and Astronomy,
Bloomberg Center, 3400 N. Charles Street, Baltimore, MD 21218 (nl@pha.jhu.edu)}

\begin{abstract}
We discuss 
{\em Far Ultraviolet Spectroscopic Explorer} (\fuse) observations of two early-type stars, DI\,1388 
($l = 291.3 \degr$, $b = -41.1\degr$)
and DGIK\,975 ($l = 287.3\degr$, $b = -36.0 \degr$), 
in the low density and low metallicity ($Z\sim0.08 Z_\odot$) 
gas of Magellanic Bridge (MB). The data have a spectral resolution of 
about 15,000 and  signal-to-noise ratios range between 10 and 30 per resolution element in the 
spectra of DI\,1388 and between 7 and 11 in the spectra of DGIK\,975. DI\,1388 is situated near the SMC,
while  DGIK\,975 is closer to the LMC, allowing us to probe the MB gas in a widely different locations.
Toward DI\,1388, the \fuse\/ observations show
molecular hydrogen,  \ion{O}{6}, and numerous other
atomic or ionic transitions in absorption, implying  
the presence of multiple gas phases in a complex arrangement. 
The relative abundance (with respect to \ion{S}{2}) pattern in the MB along the
DI\,1388 sight line is attributed to varying degrees of depletion 
onto dust similar to that of halo clouds. The N/O ratio
is near solar, much higher than N/O in damped Ly$\alpha$ systems, implying 
subsequent stellar processing to explain the origin of nitrogen in the MB. 
The diffuse molecular cloud in this direction has a low column density and low molecular fraction 
($\log N ({\rm H}_2) \approx 15.43$ dex; 
$f_{\rm H_2} \sim 10^{-5} - 10^{-4} $),
yet two excitation temperatures ($T_{01} = 94 \pm\,^{53}_{27}$ K 
and $T_{23} = 341 \pm\,^{172}_{81}$ K) are needed to fit the distribution of the different
rotational levels. Though this is not typically seen in the Galaxy,
we show that this is not uncommon in the Magellanic Clouds. 
H$_2$ is observed in both the Magellanic Stream and the MB, yet massive stars form only in the MB, 
implying significantly different physical
processes between them. In the MB some of the H$_2$ could have been
pulled out from the SMC via tidal interaction, but some also could
have formed {\em in situ} in dense clouds where star formation might have taken place.
Toward DGIK\,975, the presence of neutral, weakly and highly ionized 
species suggest that this sight line has also several complex gas phases. 
The highly ionized species of \ion{O}{6}, \ion{C}{4}, and \ion{Si}{4} toward both stars
have very broad features, indicating  that 
multiple components of hot gas at different velocities are present. 
\ion{C}{4}/\ion{O}{6} varies within the MB but \ion{C}{4}/\ion{Si}{4} is 
relatively constant for both sight lines.  Several sources (a combination of turbulent mixing layer, 
conductive heating, and cooling flows) may be contributing to the production of the highly
ionized gas in the MB.
Finally, this study has confirmed previous results
that the high-velocity cloud HVC $291.5-41.2+80$ is mainly ionized composed of weakly 
and highly ions. The high ion ratios are consistent with a 
radiatively cooling gas in a fountain flow model.
\end{abstract}


\section{Introduction}
The Magellanic system is composed of two small irregular
galaxies, the Large (LMC) and Small (SMC) Magellanic Clouds, 
in orbit around the Galaxy. Tidal interactions between these
galaxies have produced several 
high velocity  gas complexes connected to the Clouds \citep[for a recent study, see][]{putman00}; 
namely the Magellanic Bridge (MB), the Magellanic Stream, and the leading Arm. 
In particular for this study, the Magellanic Bridge is a $10\degr$ region of tenuous gas 
linking the body 
of the SMC to an extended arm of the LMC.  
The formation mechanism responsible for this feature
remains unclear, but it is generally agreed that the MB
was formed via a tidal encounter between the SMC and LMC. \citet{gardiner}
have produced models that can reproduce simultaneously both the MB and 
the Magellanic Stream, and find that the MB was most likely pulled from
the wing of the SMC 200 Myr ago during a close encounter between the 
two Clouds. 
However, its low metallicity 
\citep[$Z \approx 0.08 Z_\odot$ based on C, N, O, Mg, and Si][]{rol99}
does not reflect the current SMC metallicity \citep[$Z \approx 0.25 Z_\odot$, e.g.,][]{russell},
suggesting that the MB gas could be formed from a mixture of SMC gas and an unenriched
component or could be much older than the age predicted from theoretical models.

The MB contains early-type stars \citep{hambly,demers98,rol99}. During their
main-sequence lifetimes (as short as 20 Myr), these stars
could not possibly migrate from the SMC since they do not 
exhibit peculiar velocities sufficient to explain their
motion over the large distances they would need to cover. Therefore the MB provides 
the most metal poor gas in our neighborhood to investigate not
only the gas, but also to observationally constrain star-formation
in low metal and tenuous gas environment. Furthermore, 
it provides an unique opportunity to study in detail the 
gas resulting from tidal interactions.  In the Galaxy, star formation
generally occurs in dense molecular clouds \citep[e.g.,][]{evans99}.
No direct evidence of molecular clouds have been discovered yet \citep{smoker00}, although
\citet{kob99} found cold atomic clouds, suggesting molecular condensations. 

\citet{lehner01a} recently reported the results 
of a program to investigate the chemical composition and abundance pattern 
of the MB gas toward an early-type star, DI\,1388, with the {\em Hubble Space Telescope} {\em (HST)}\/
and the Space Telescope Imaging Spectrograph (STIS). The combination 
of high spectral resolution and high sensitivity in the ultraviolet
bandpass ($1150-1730$ \AA) made it feasible to 
investigate the chemical composition and the physical conditions within the MB
and revealed complex gas phases with neutral gas, weakly and highly ionized gas
along the sight line studied. Yet only at lower wavelengths than the STIS bandpass, 
in the 905--1187 \AA\ wavelength range, are 
the strong hydrogen molecular lines accessible. 
Other strong atomic and ionic resonance lines are as well only limited in this 
bandpass, such as the \ion{O}{6} doublet, a powerful diagnostic of collisionally 
ionized gas. \ion{O}{6} is unlikely to be produced by photoionization alone from starlight given
that photons with $h\nu \ge 114$ eV are needed to convert \ion{O}{5} to \ion{O}{6}.
Such a bandpass is accessible now with the {\em Far Ultraviolet Spectroscopic Explorer} ({\fuse})
and therefore follow-up observations of the low metallicity, tenuous gas of the MB 
were obtained with this observatory. 

In this article, we report on {\fuse}\/ observations of two early-type stars in the MB,  DI\,1388  
which is situated approximately mid-way between the SMC and LMC  and DGIK\,975, which lies at the western 
end of the Bridge near to the LMC halo. The {\fuse}\/ spectra allow us for 
the first time to make a sensitive and direct search for molecules in the MB. They 
provide a means to study the collisional gas seen in \ion{O}{6} absorption and to compare
with recent \ion{O}{6} surveys in the SMC and LMC \citep{hoopes02,howk02} and in the Galactic
halo \citep{savage00}. They also provide 
a quantitative estimate of the quantities of neutral and weakly ionized gas in the MB and in the 
ionized high-velocity cloud HVC $291.5-41.2+80$ \citep{lehner01b}.

\section{Observations and Data Processing}
We obtained {\fuse}\/ spectra of two early-type stars, DI\,1388 
($l = 291.3 \degr$, $b = -41.1\degr$)
and DGIK\,975 ($l = 287.3\degr$, $b = -36.0 \degr$), 
in the MB on 2001 August 15 and August 24, respectively.
A summary of their stellar properties and location is given in Table~\ref{t0}.
The data are cataloged in the MAST archive under the 
identifications P2410101 and P2410201. Total exposure times of 
17,460 seconds for DI\,1388 and
23,520 seconds for DGIK\,975 were obtained.
All the data were obtained 
in time-tag mode through the $30 \arcsec \times 
30 \arcsec$ aperture. 

The {\fuse}\/ instrument consists of four channels: two optimized for the short
wavelengths (SiC\,1 and SiC\,2; 905--1100 \AA) and two optimized 
for longer wavelengths (LiF\,1 and LiF\,2; 1000--1187 \AA). There is,
however, overlap between the different channels, and, generally,
an atomic or a molecular transition appears in at least two different
channels. More complete descriptions of the 
design and performance of the {\fuse}\/ spectrograph are given 
by \citet{moos00} and \citet{sahnow00}, respectively.
To maintain optimal spectral resolution and
information on the fixed-pattern noise, the individual segments 
were not co-added together. For both 
sight lines, all the four channels were reasonably co-aligned throughout
the observation, but for DGIK\,975, detector 2 was shutdown
for most of the observation resulting in the loss of SiC\,2 and
LiF\,2 data (note as well that 3/14 exposures in SiC\,1 were lost). Therefore
only the LiF\,1 channel is used in this work for the DGIK\,975 sight line.

Standard processing with the current version of the calibration
pipeline software ({\sc calfuse} v2.0.5) was used to extract
and calibrate the spectra. The software screened the data
for valid photon events, removed burst events, corrected
for geometrical distortions, spectral motions, satellite orbital
motions, and detector background noise, and finally applied
flux and wavelength calibrations.
The extracted spectra associated with the separate exposures 
were aligned by cross-correlating the positions of
strong interstellar lines, and then co-added. The co-added spectra
were rebinned by 4 pixels ($\sim 27$ m\AA) since the extracted data are oversampled. 
Scattered light and detector backgrounds are negligible at the flux
levels of the absorption lines considered in this work.
The spectra have a nominal spectral resolution of
$\sim$$15,000$ ($\approx 20$ \km; 2 pixels FWHM). 
For DI\,1388, the spectra have a 
signal-to-noise (S/N) ratio in the continuum between 10 and 30 per resolution
element; and for DGIK\,975 between 7 and 12 per resolution
element. In Tables~\ref{t1} and \ref{t2}, the S/N levels are given 
for each studied transitions and were determined by examining the 
dispersion about low-order polynomial fit to the continuum
in the vicinity of the specified absorption lines.
The relative wavelength calibration is  accurate to about $\pm 5$ \km\ but 
can vary by 10--15 \km\ over small wavelength intervals. 
The zero point of the wavelength scale was set 
by comparing absorption at {\fuse}\/ wavelengths to longer UV or optical wavelengths.
The MB component is present at $v_{\rm LSR} \sim 200$ \km\ toward DI\,1388, and 
at $v_{\rm LSR} \sim 170$ \km\ toward DGIK\,975. 

\begin{figure*}[!th]
\begin{center}
\includegraphics[width=18truecm]{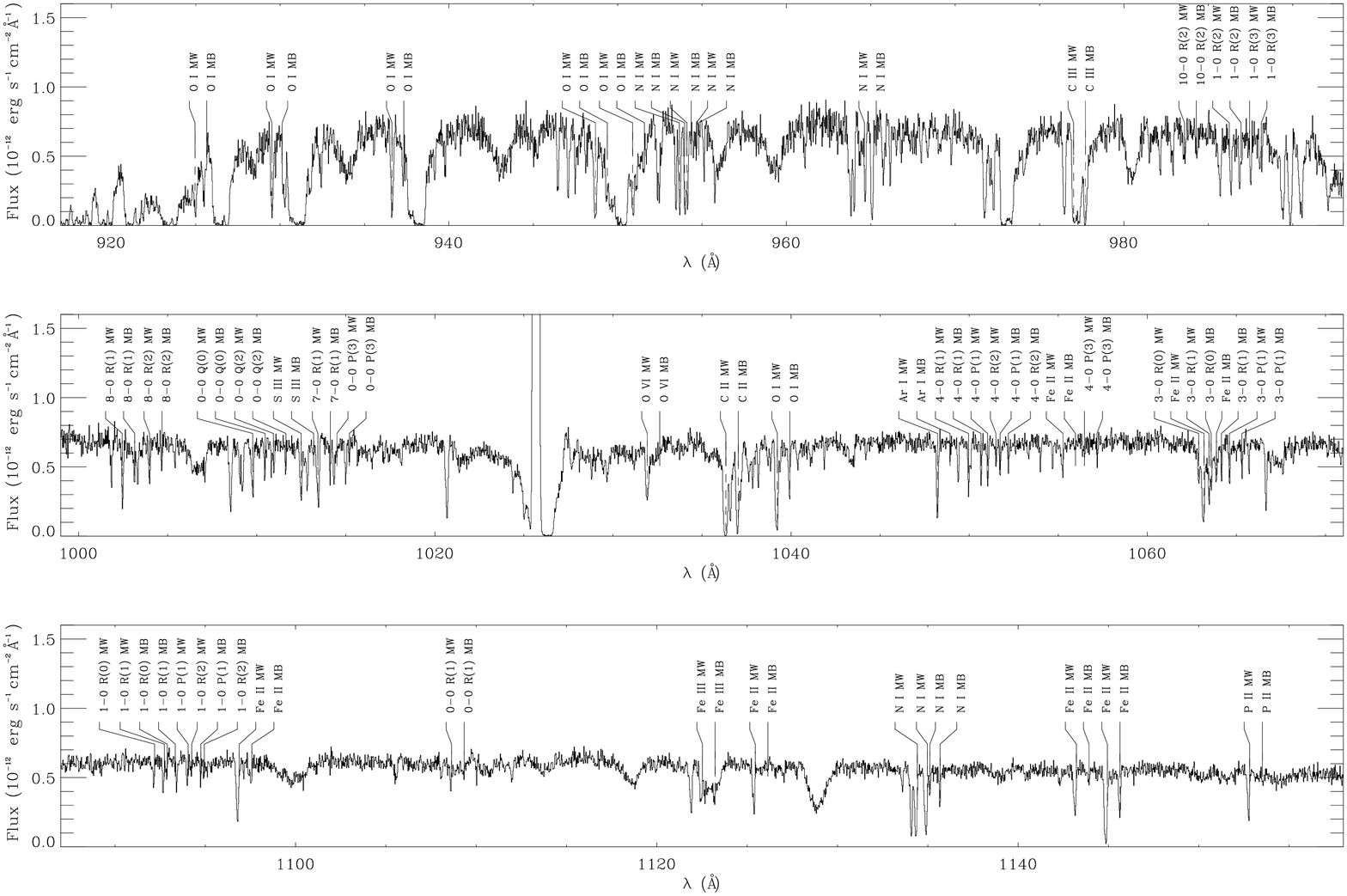}
\caption{Far-ultraviolet spectra of DI\,1388 with identification of 
some interstellar atomic and molecular lines, 
arising in the Galaxy (MW) and the Magellanic Bridge (MB).} 
\label{fig0}
\end{center}
\end{figure*}

We illustrate in Figure~\ref{fig0} the reduced \fuse\/ spectra of DI\,1388 for different detector segments. 
The data reveal numerous stellar and interstellar 
atomic and molecular absorption lines. Interstellar lines are identified
when both the Galactic and MB components are present. The \ion{H}{1} Lyman-series 
can be easily identified. Broader features are stellar lines. Because
the {\fuse}\/  bandpass is very rich in interstellar lines, a careful 
check was made to ensure that no other line was blended with the MB component.
We note that while the airglow lines are relatively strong (see Figure~\ref{fig0},
especially \ion{H}{1} and \ion{O}{1}), they do not affect the MB measurements
as those lines are shifted by $\ga+150$ \km.

\section{Analysis} \label{analysis}
\begin{figure*}[!th]
\begin{center}
\includegraphics[width=16truecm]{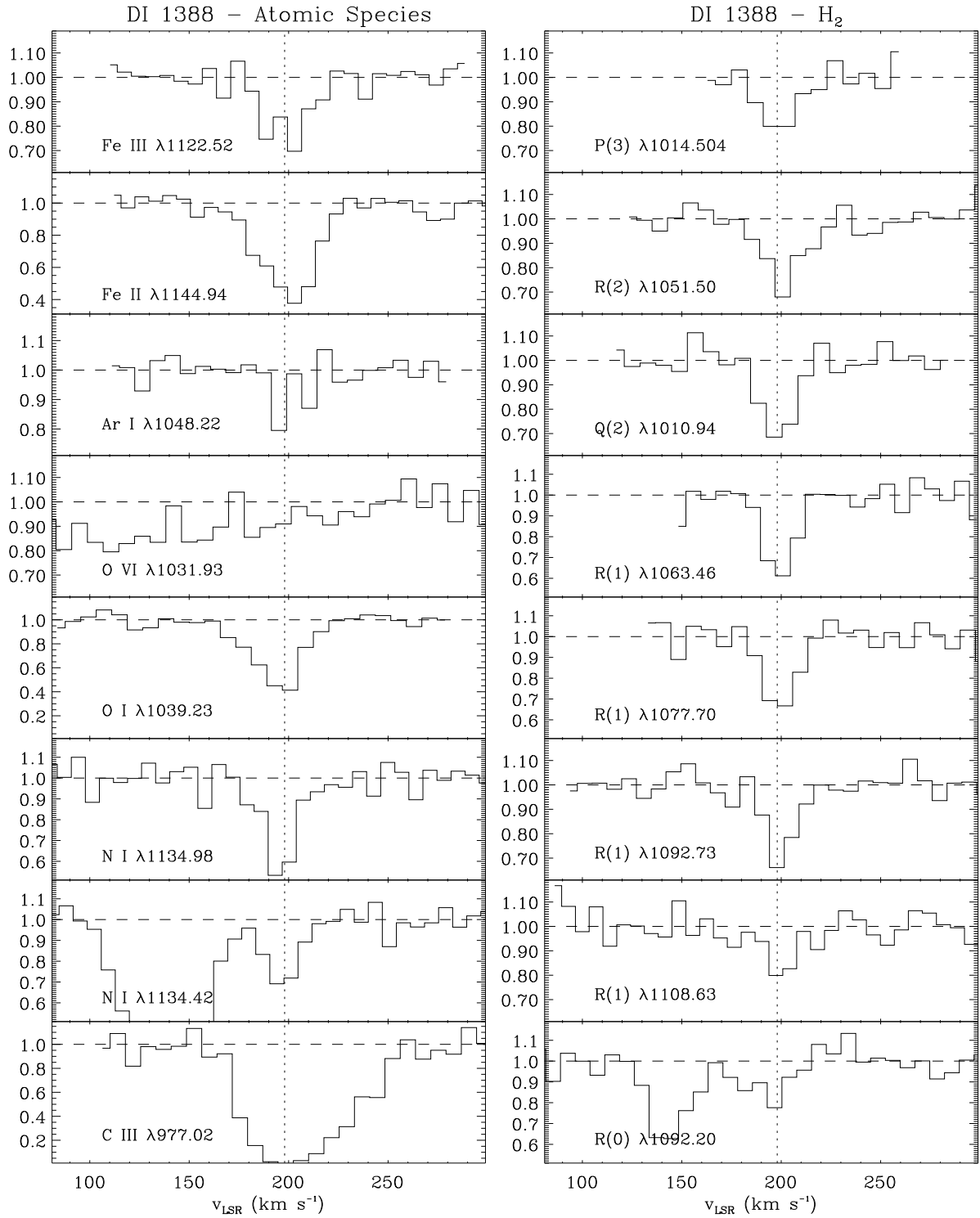}
\caption{Normalized profiles for selected Magellanic Bridge 
interstellar lines ($v_{\rm LSR} \sim 200$ \km) in the  DI\,1388 spectrum. } 
\label{fig1}
\end{center}
\end{figure*}

\begin{figure*}[!th]
\begin{center}
\includegraphics[width=9truecm]{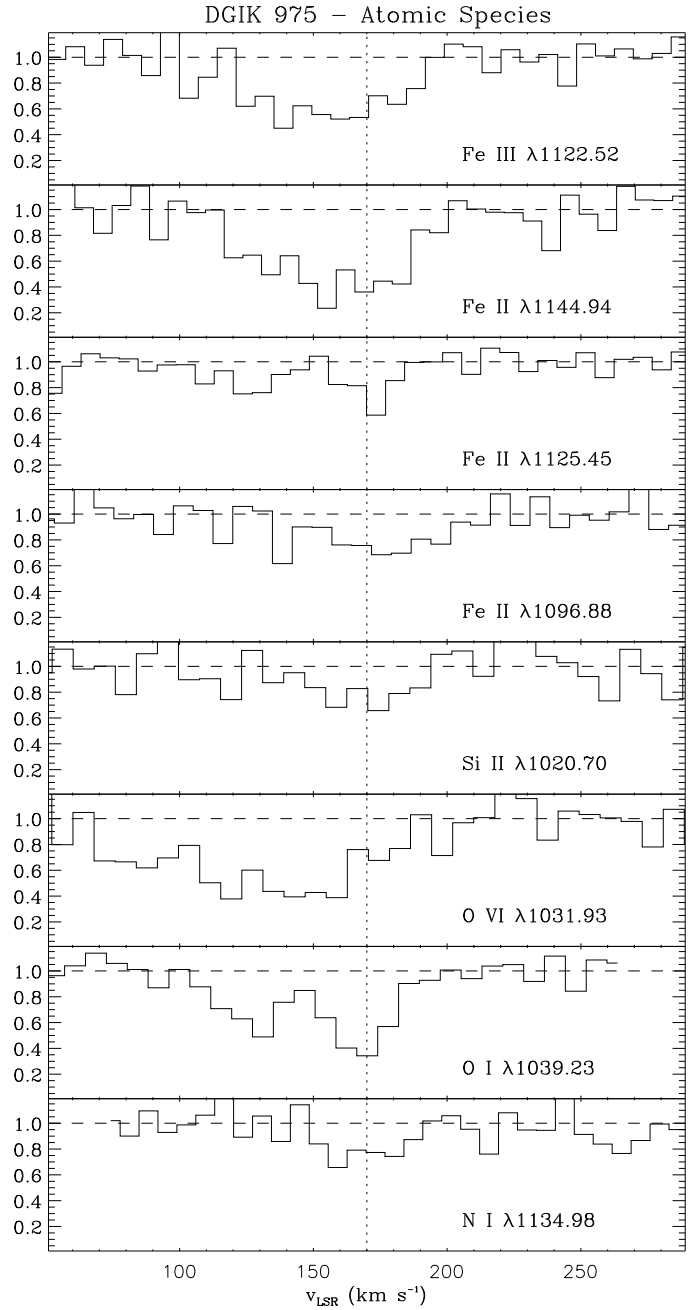}
\caption{Normalized profiles for selected Magellanic Bridge 
interstellar lines ($v_{\rm LSR} \sim 170$ \km) in the  DGIK\,975 spectrum. } 
\label{fig2}
\end{center}
\end{figure*}

The stellar continua were simple enough near the intersellar lines in the \fuse\/ bandpass
to be fitted with low-order ($\le 4$) Legendre polynomials.
A selection of normalized profiles is shown in Figures~\ref{fig1}
and \ref{fig2} for the DI\,1388 and DGIK\,975 sight lines, respectively.
Species detected in the {\fuse}\/ spectra in absorption are: (1) For DI\,1388, 
\ion{C}{2}, \ion{C}{3}, \ion{N}{1}, \ion{N}{2},  \ion{N}{3}, \ion{O}{1}, 
\ion{O}{6}, \ion{P}{2}, \ion{S}{3}, \ion{Ar}{1}, \ion{Fe}{2}, \ion{Fe}{3}, and
H$_2$. ; (2) For DGIK\,975, \ion{C}{2}, \ion{N}{1}, \ion{N}{2}, \ion{O}{1},
\ion{O}{6}, \ion{Si}{2}, \ion{Fe}{2}, and \ion{Fe}{3}. The absorption lines are stronger 
toward DGIK\,975, and therefore more likely to be saturated.
Moreover, the lower S/N of the  DGIK\,975
spectra hinders precise measurements toward this 
sight line and also limits the possibility of detecting weak features. 

The equivalent widths were measured by directly integrating the intensity
and are listed in Table~\ref{t1} and \ref{t2}. 
The equivalent width 
measurements and errors were derived following the method described by 
\citet{sembach92}. 

To obtain the column density, two methods were used: 

(1) The apparent
optical depth (AOD) method \citep{savage}  was used when only one or two lines
of the same atomic species were available. 
The absorption profiles were converted into apparent
optical depths per unit velocity, $\tau_a(v) = \ln[I_{\rm c}/I_{\rm obs}(v)]$, 
where $I_{\rm c}$, $I_{\rm obs}$ are the intensity without and with the absorption,
respectively.  $\tau_a(v)$ is related
to the apparent column densities per unit velocity, $N_a(v)$ 
(cm$^{-2}$ (\km)$^{-1}$) through the relation 
$ N_a(v) = 3.768 \times 10^{14} \tau_a(v)/[f \lambda(\rm\AA)]$.
The integrated apparent column density is equivalent to the 
true integrated column density if no unresolved saturated structure is present.
The results are summarized in Tables~\ref{t1} (DI\,1388)
and \ref{t2} (DGIK\,975). The results for the \ion{O}{6}  absorption profile
and other highly ionized species are discussed in \S~\ref{hot} (DI\,1388)
and \S~\ref{dgik975} (DGIK\,975).

\begin{figure*}[!th]
\begin{center}
\includegraphics[width=12truecm]{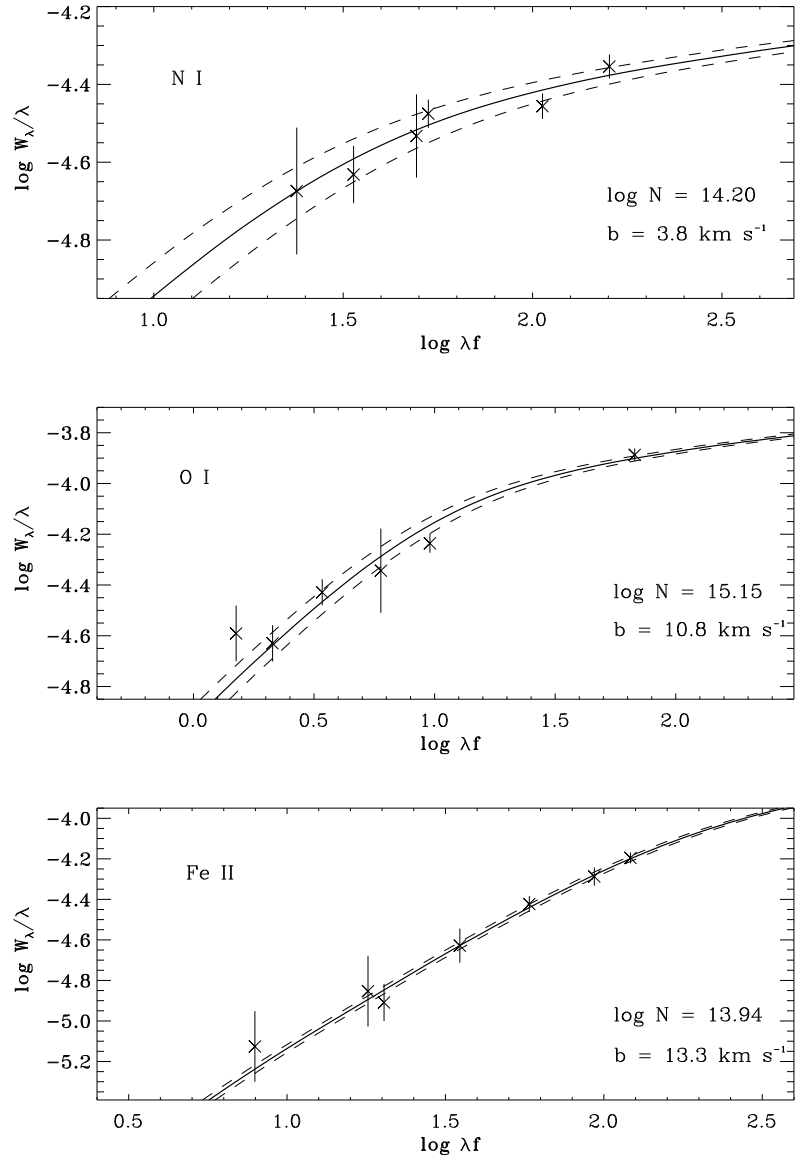}
\caption{Single-component curve of growth for the \ion{N}{1}, \ion{O}{1}, and 
\ion{Fe}{2} lines identified in the spectrum of DI\,1388. The dashed lines 
reflect the $1\sigma$ error on the column density.
} 
\label{fig3}
\end{center}
\end{figure*}

\begin{figure*}[!th]
\begin{center}
\includegraphics[width=18.truecm]{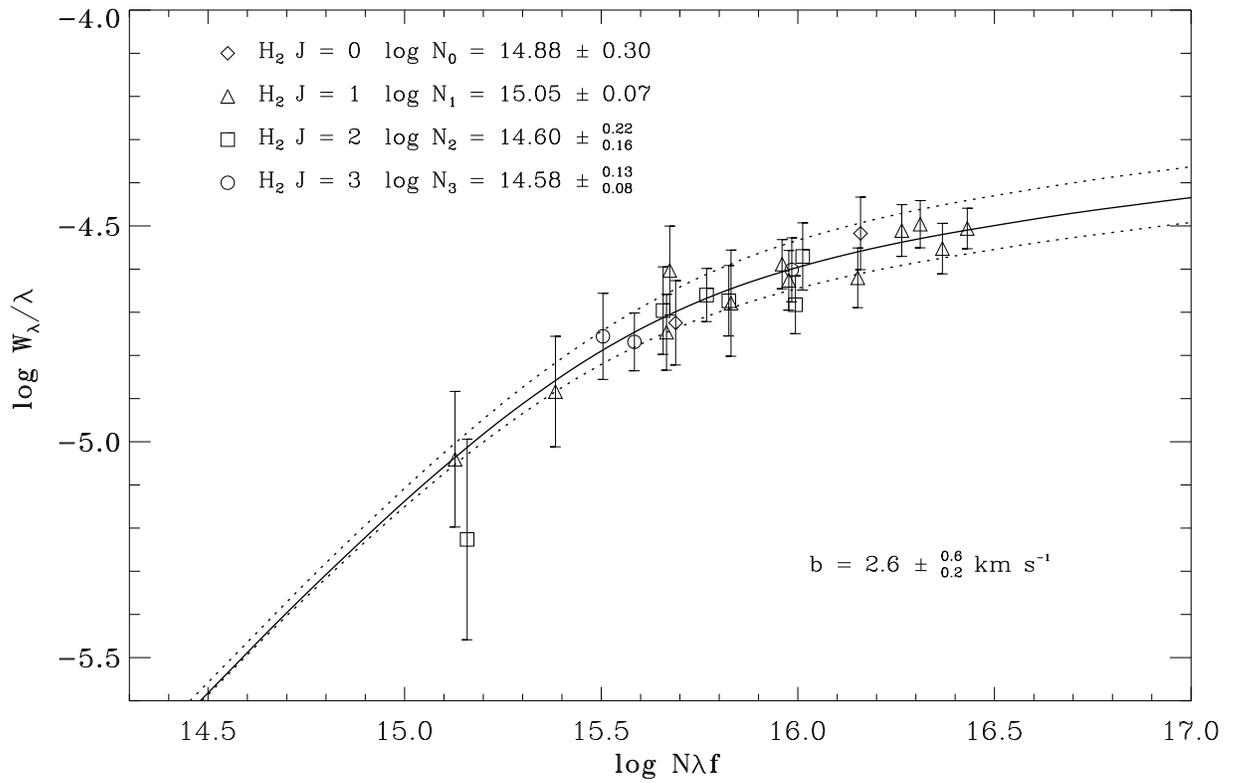}
\caption{Single-component curve of growth for the H$_2$ lines identified in the spectrum
of DI\,1388. The dotted lines are the curve of growth for $b$-values of 2.4 and 3.2 \km.} 
\label{fig4}
\end{center}
\end{figure*}

(2) A curve-of-growth (COG) analysis was undertaken 
when several lines were present
(\ion{N}{1}, \ion{O}{1}, \ion{Fe}{2} and H$_2$ toward DI\,1388) by
using measured equivalent widths. 
Additional lines for the atomic species were used from the 
measurements in the {\em HST}/STIS spectra \citep[see, ][]{lehner01a}.
A single component Gaussian (Maxwellian)
curve of growth was constructed in which the Doppler parameter $b$ and 
the column density were varied to minimize the $\chi^2$ between the observed equivalent
widths and a model curve of growth. For DI\,1388, the results are summarized
in Figures~\ref{fig3} (atomic species) and \ref{fig4} (H$_2$). For H$_2$,
a COG was constructed for the lines within each rotational level ($J= 0,...,3$; no detectable
absorption is present in levels $J \ge 4$), and this resulted
in the COG presented in Figure~\ref{fig4}, 
describing simultaneously all the rotational levels
with a Doppler parameter of $b = 2.6 \pm \,^{0.6}_{0.2}$ \km. The $J = 0$ level is poorly constrained
as only two transitions were available. 

Good agreement 
is found between the AOD and COG methods for \ion{Fe}{2}, principally because
the transitions for this ion are not saturated. The transitions of the neutral
species lie near or on the knee of the COG where saturation effects
arise. The $b$-values are comparable to the total $b$-values
obtained by \citet{lehner01a} with the higher spectral resolution of the 
{\em HST}/STIS data using a profile
fitting method; in particular, the same pattern is observed, i.e. a
smaller $b$-value for \ion{N}{1} compared to \ion{O}{1} and \ion{Fe}{2},
but a similar $b$-value for \ion{O}{1} and \ion{Fe}{2}. 
For DGIK\,975, 
the \ion{Fe}{2} column density derived with the COG method
is in the range given by the AOD method, but because the lines are much stronger,
the S/N is low, and only four transitions are available, 
the $b$-value is not constrained. Optical \ion{Ca}{2} K spectra of 
DGIK\,975 with similar quality to those described in \citet{lehner01a} toward
DI\,1388 (spectral resolution of about 7.1 \km) do not show any
evidence of multiple component structure. However, \ion{Ca}{2} K cannot 
{\em a priori} be used to model \ion{Fe}{2} component
structure in diffuse clouds, as it traces mainly cold neutral clouds, while \ion{Fe}{2} can be
found in both neutral and ionized clouds.\footnote{D.E. Welty (private communication, 2002) has
noted, however, that comparisons of high spectral resolution profiles for \ion{Ca}{2} 
and \ion{Fe}{2} can show surprising similarity in many cases.}
Note as well that finer structure seen in higher 
resolution spectra of Galactic HVCs \citep[e.g.,][]{lehner99} suggests that the true component 
structure in the MB gas toward both sight lines may be more complex than is revealed
by 6.5 \km\ or lower spectral resolution.

Since we have information on \ion{N}{1} and \ion{N}{2} toward DI\,1388, it 
is interesting to estimate the amount of interstellar \ion{N}{3}. 
\ion{N}{3} $\lambda$989.799 is severely blended with \ion{Si}{2} $\lambda$989.873. 
Using the equivalent measurements available in the STIS bandpass for \ion{Si}{2}
\citep{lehner01a}, we constructed a COG and estimated the equivalent
width for \ion{Si}{2} $\lambda$989.873 as being $\sim$105 m\AA. Knowing
the total equivalent width of this feature and assuming that neither lines
suffer from saturation effect, we infer that the \ion{N}{3} absorption 
feature has an equivalent width of $\sim$31 m\AA, and 
with the assumption that it lies on the linear part of the COG, 
its column density is $\sim$13.5 dex.

The $3\sigma$ upper limits for the equivalent widths in Table~\ref{t1}
are defined as $W_{\rm min} = 3 \sigma\, \delta\lambda$, where 
$\sigma$ is the inverse of continuum S/N ratio and $\delta \lambda \approx 0.1$ \AA,
which is approximately the average FWHM obtained from the resolved features.
The corresponding  $3\sigma$ upper limits on the column density
are obtained from the corresponding equivalent width limits and the assumption 
of a linear curve of growth.

We adopted wavelengths and oscillator strengths from 
the Morton (2000, private communication) atomic data compilation. 
This compilation
is similar to the \citet{morton91} compilation with a few minor updates to the atomic
parameters for lines of interest in this study.
For the \ion{Fe}{2} lines, the new oscillator strengths derived by \citet{howk00}
were adopted. We note that there is a new laboratory measurement of the $f$-value 
for \ion{Fe}{2} $\lambda$1144.9
\citep[$f = 0.083 \pm 0.006$,][]{wiese02}, about 30\% lower than the 
experimental value of \citet{howk00}. We note, however, that the 
value of Howk et al. gives a column density in good agreement with 
the measurement from the \ion{Fe}{2} $\lambda$1608.4 line. Moreover, 
\citet{wiese02} note that additional study is necessary for
the other transitions available in the \fuse\/ bandpass, rather
than a simple rescaling, as they are further down on the COG
than the \ion{Fe}{2} $\lambda$1144.9 line. As we use all the transitions 
available in \fuse\/ spectrum, we employ the $f$-value of \citet{howk00}
and we note that even if $f$-values are lower, 
the column density would only change by about $+0.05$ dex, and
our conclusions would remain the same. 
The H$_2$ wavelengths are from \citet{abgrall93a,abgrall93b}, while
the H$_2$ $f$-values were calculated from the emission probabilities given by those
authors.

\section{Magellanic Bridge gas toward DI\,1388}
\subsection{Molecular hydrogen}\label{h2}
Until recently, searches for molecular gas in the MB
were restricted to observations of CO \citep{smoker00}, but none was
found, mainly due to a lack of sensitivity. 
The most abundant molecule in the Universe, molecular hydrogen 
(H$_2$) has only been recently accessible with the launch of {\em FUSE}\/
for faint targets observed in the MB. {\em FUSE}\/ 
observations reveal that 92\% and 52\% of the SMC and LMC sources, respectively,
exhibit H$_2$ absorption lines \citep{tumlinson02}. Two observations
toward the Magellanic Stream also reveal H$_2$ \citep{sembach01,richter01}. 
While the S/N is too low in the spectrum of DGIK\,975 to allow 
any conclusions, the spectrum of DI\,1388
exhibits  about 30 absorption lines from the Lyman and Werner bands of H$_2$
(see Table~\ref{t1}). In all known environments in our Galaxy, star formation occurs within
molecular clouds. The detection of H$_2$ toward DI\,1388 is therefore
the first direct evidence of a possible star formation signature
in the very poor metal environment of the MB. 

From Table~\ref{t1} and Figure~\ref{fig4}, 
the total column density of H$_2$ is $\log N ({\rm H}_2) \approx 15.43$ dex.
For $J \ge 4$, $\log N_J({\rm H}_2) <  14.0$ dex. 
The \ion{H}{1} column density is not well known \citep{lehner01a}: the \ion{H}{1} emission 
column density is 20.50 dex (Putman 2000, private communication; see Lehner et al.
2001a). In theory, the {\fuse}\/ observations could help to 
constrain the \ion{H}{1} absorption column density because 
numerous \ion{H}{1} lines of the Lyman series with different oscillator strengths are accessible. 
Moreover, at $\lambda < 926.2$ \AA, the MB component can be separated 
from the lower velocity components. But as displayed in Figure~\ref{fig0}, 
the continuum position is uncertain at these wavelengths. The stellar 
contribution is also not known due to the difficulty to model 
this star because its large $v\sin i$. Therefore, the 
\ion{H}{1} absorption column density cannot be directly derived. 
An alternative is to use the \ion{O}{1} column density. 
\ion{O}{1} is an excellent tracer of neutral gas as
its ionization potential and charge exchange reactions with hydrogen
ensure that the ionization of \ion{H}{1} and \ion{O}{1} are strongly coupled. 
Using the \ion{O}{1} column density, and
assuming the interstellar O abundance\footnote{Meyer et al.'s  value 
($ {\rm O/H} = (3.19 \pm 0.14)\times 10^{-4}$) was corrected for
the recommended oscillator strength of \ion{O}{1} $\lambda$1356 ($f= 1.16 \times 10^{-6}$
instead of $f= 1.25 \times 10^{-6}$) by 
\citet{welty99b}.} found by \citet{meyer98}
$ {\rm O/H} = (3.43 \pm 0.15)\times 10^{-4}$, and a general deficiency 
in metal of the MB gas of $-1.1 $ dex from the stellar study \citep{rol99}, we obtain
$\log N$(\ion{H}{1}$)\approx 19.72$ dex. The
difference could be due to the large beam of the 
\ion{H}{1} emission data and to the position (in depth along the sight line) 
of DI\,1388 in the MB. 

Using the \ion{H}{1} value derived from \ion{O}{1}, the fractional 
abundance of H$_2$ is $f_{\rm H_2} \equiv 2 N ({\rm H}_2)/[N$(\ion{H}{1}$) + 2 N ({\rm H}_2)] \sim
1 \times 10^{-4}$, while the \ion{H}{1} emission column density suggests
$f_{\rm H_2} > 2 \times 10^{-5}$. This value is comparable to those seen along
Galactic sight lines \citep{savage77}
and Magellanic Clouds sight lines \citep{tumlinson02} when E$(\bv) \la 0.1$ and 
$\log [N$(\ion{H}{1}$) + 2 N ({\rm H}_2)] \la 20.5$ dex. However, \citet{tumlinson02}
noted that a major difference between the Galactic and Magellanic Clouds 
samples is the presence of several sight lines with $f_{\rm H_2} \ga 10^{-4}$
when  E$(\bv) \la 0.08$. We also note that for similar H$_2$ column densities
in the SMC and LMC, molecular fractions similar to that derived here are observed.
For the MB, and in some cases for the SMC and LMC (see Figure~\ref{fig7} in the
Appendix), $f_{\rm H_2}$ remains uncertain due to \ion{H}{1} column density uncertainties.

\begin{figure*}[!th]
\begin{center}
\includegraphics[width=18.truecm]{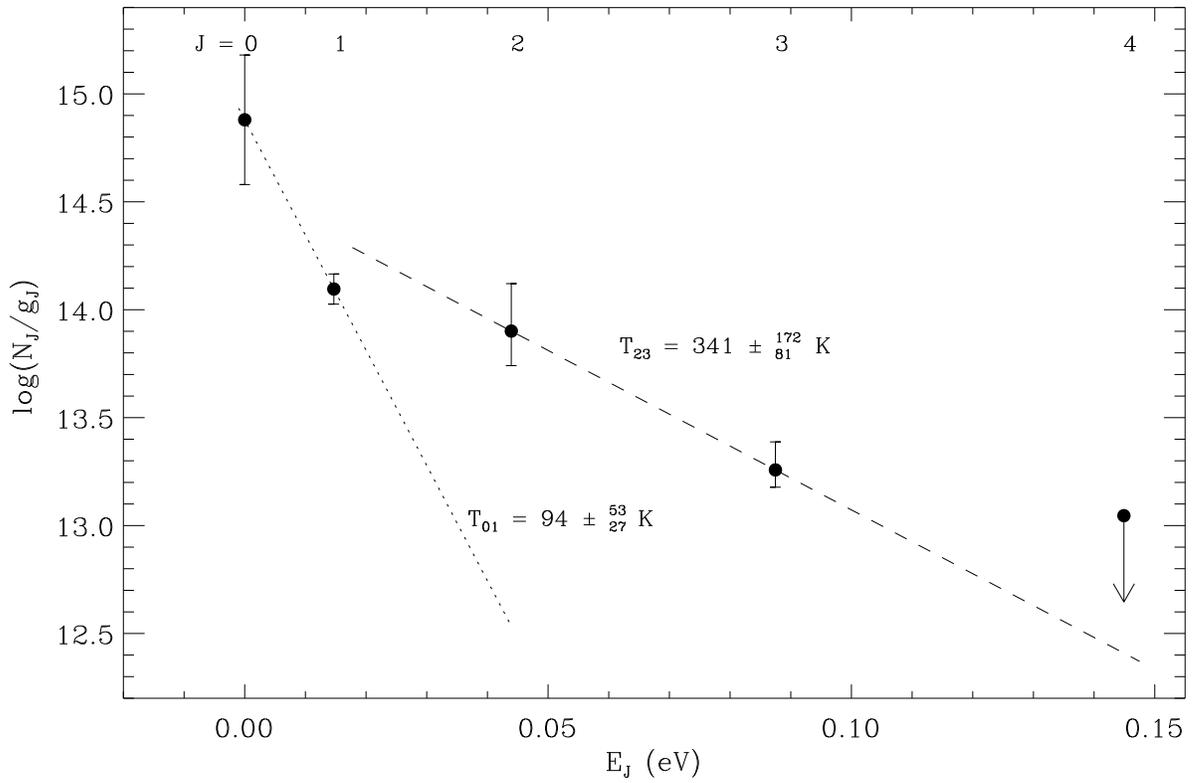}
\caption{H$_2$ column density ($N_J$) divided by statistical
  weight ($g_J$) as a function of the excitation energy ($E_J$) for rotational
  levels $J = 0$--4 in the MB toward DI\,1388. The data point for $J = 4$ is 
  a 3$\sigma$ upper limit. All other errors are 1$\sigma$ estimates.} 
\label{fig5}
\end{center}
\end{figure*}

The population of the lower rotational states of molecular hydrogen 
is determined by collisional excitation with a Boltzmann
distribution, so that the excitation temperature $T_{ij}$
is given by \citep[e.g.,][]{spitzer74}
\begin{equation}\label{eq1}
T_{ij} = \frac{E_j-E_i}{k \ln[\frac{g_j}{g_i}\frac{N_i}{N_j}]} \,.
\end{equation}
We find that $T_{01} = 94 \pm\,^{53}_{27}$ K and $T_{23} = 341 \pm\,^{172}_{81}$ K.
In Figure~\ref{fig5}, the excitation 
temperature diagram is presented for H$_2$ arising in ground-state rotational levels
$J \le 4$. The interpretation of this 
diagram is complicated by the large uncertainty in the column density
of the $J = 0$ level.  
In their survey in the Galaxy, \citet{spitzer74} found
that when $N(J=0) \la 10^{15}$ cm$^{-2}$, a single excitation temperature
generally fits all the observed $N(J)$ within their estimated errors.
Toward DI\,1388 $N(J=0) \approx 10^{14.88}$ cm$^{-2}$, but 
we believe that the estimated error
of this level is conservative and therefore a single temperature
does not describe the distribution of the data points within the error 
bars. In the Appendix, we show by using the LMC and SMC sample
that for low H$_2$ column density, a single excitation temperature
does not {\em generally} fit the distribution of the data points,
as observed in the MB, when $N(J=0) \la 10^{15}$ cm$^{-2}$.

The $J=0,1$ temperature, though uncertain, is in 
general agreement with what is derived in diffuse molecular clouds in the Magellanic Clouds 
\citep[$\langle T_{01} \rangle = 115$ K for all their sight lines;][]{tumlinson02},
the Galactic halo \citep[$\langle T_{01} \rangle = 107 \pm 17$ K;][]{shull00}, and in 
the Galactic disk \citep[$\langle T_{01} \rangle = 77 \pm 17$ K;][]{savage77}. 
The absolute value of  $T_{01}$ suggests a closer similarity to the Clouds
and the Galactic halo than the Galactic disk, yet the large error on  $T_{01}$ prevents us from drawing
definitive conclusion. While $T_{01}$ may trace directly the kinetic 
temperature in dense molecular clouds, in low column diffuse clouds
the relationship between the excitation temperature of the first two levels
and the kinetic temperature is not well understood \citep[e.g.;][]{tumlinson02}.

\subsection{Neutral and partially ionized gas}
\citet{lehner01a} found two MB interstellar clouds toward DI\,1388
at 179 and 198 \km\ with the 6.5 \km\ spectral resolution of the 
{\em HST}/STIS.
The 198 \km\ cloud is mainly neutral with a hot component (see \S~\ref{hot}); 
while the 179 \km\ cloud is warmer, low density gas, and partially ionized.
The {\fuse}\/ bandpass gives access to a large number of ions not accessible
with STIS, though at lower resolution, which complicates the interpretation 
of the data.  In particular the separation between the two clouds is about
the resolution of {\fuse}\/ and therefore theses clouds  
cannot be resolved with {\fuse}. 
However, some properties of this sight line can still be derived,
since for the neutral and singly ionized species we know from
\citet{lehner01a} that the mainly neutral 
gas at 198 \km\ is about 2--3 times stronger in column density than the ionized
cloud at 179 \km. 

The detection of H$_2$ at $\sim$200 \km\ suggests that it is associated 
with the neutral cloud observed in the atomic and ionic species. While 
the excitation temperature $T_{01}$ is relatively low, it is not known
how it is related to the kinetic temperature in diffuse clouds. However, the
$b$-value obtained from the COG analysis for H$_2$ gives an upper limit to the 
kinetic temperature of $ \la 800$ K. 
We note that both the COG and profile fitting analysis give a lower $b$-value
for \ion{N}{1} compared to \ion{O}{1}, suggesting that the mostly neutral cloud
is probably composed of more than one cloud. This is also supported by the 
fact that the profile fitting to \ion{O}{1} $\lambda$1302 indicates that some 
unresolved structures remain \citep[see Figure 2 in][]{lehner01a}. 

For the MB gas we find, when considering the 
total column density, $N($\ion{Fe}{2}$)/N($\ion{Fe}{3}$) \approx 1.9$,  
$N($\ion{S}{2}$)/N($\ion{S}{3}$) \approx 4.6$. \citet{sembach00}
have produced models of low density photoionized gas, and these ratios
compare with a combination of their composite and ``$x_{\rm edge} =0.95$''
models. Such models typically simulate regions with significant fraction
of neutral hydrogen or ionized interfaces of neutral clouds.
Considering only
the ionized component for \ion{S}{2} and \ion{Fe}{2} \citep[see,][]{lehner01a}
and assuming that \ion{S}{3} and \ion{Fe}{3} is mostly present in the
ionized gas, the ratios are $N($\ion{Fe}{2}$)/N($\ion{Fe}{3}$) \approx 0.5$,  
$N($\ion{S}{2}$)/N($\ion{S}{3}$) \approx 1.2$. In this situation, the 
\ion{S}{2}/\ion{S}{3} ratio is comparable to a low density \ion{H}{2} region
with a ionization parameter $\log q = -4$ 
\citep[$\log q \ga -1$ for typical bright high density \ion{H}{2} regions, ][]{sembach00}
and ``$x_{\rm edge} =0.10$'' (i.e. fully ionized material); while \ion{Fe}{2}/\ion{Fe}{3}
is, for a low density \ion{H}{2} region,
a combination of ``$x_{\rm edge} =0.10$'' and ``$x_{\rm edge} =0.95$'' with $\log q = -4$.
Because the ionization potential of \ion{Fe}{3} is 16.18 eV smaller than
the ionization potential of \ion{S}{3}, 23.33 eV, this suggests that a fraction
of \ion{Fe}{3} could remain in the neutral gas, while \ion{S}{3} is mostly in the ionized cloud.
We note that the Sembach et al.'s models were simulated with Galactic abundances.
Metallicity effects should be negligible on ionization fractions, as long as the 
relative abundances are similar (see \S~\ref{dep}).

The determination that $N($\ion{N}{1}$)/N($\ion{N}{2}$) \la 1.0$ shows that 
N has a deficiency of its neutral form.
Because Ar (and N) is not significantly depleted 
onto dust grain \citep{sofia98}, the large deficiency of \ion{Ar}{1}
in the MB (see Table~\ref{t3}) is certainly due to photoionization. Ar is more 
easily photoionized than H because its photoionization cross section is about
10 times that of H. \ion{N}{1} has a photoionization cross section
larger than that of \ion{H}{1} and a charge-exchange
rate coefficient for \ion{N}{2} with \ion{H}{1} only twice the recombination coefficient,
so that N may show a deficiency of its neutral form, similar to what is observed for Ar
\citep{jenkins00}. We note also that \ion{N}{2}
is the dominant ion along this sight line as $N($\ion{N}{3}$)/N($\ion{N}{2}$) \la 0.2$.
Because the $b$-values for 
\ion{N}{1} and H$_2$ are similar, it could be that \ion{N}{1}, H$_2$, 
and the remaining \ion{Ar}{1} are in the colder part of the cloud, while
some of \ion{O}{1} is partly in the cold part but as well in another mostly
neutral layer of gas, where sufficiently energetic photons have ionized
N and Ar (assuming the different layers/clouds are spatially related).
 
\subsection{Highly ionized species}\label{hot}
\citet{lehner01a} found \ion{Si}{4} and \ion{C}{4} 
interstellar absorptions in the spectrum of DI\,1388.
\ion{O}{6} is the most important ion for the study of hot gas
and in particular the ratio of \ion{C}{4}/\ion{O}{6}
is an important key to differentiate
between different theoretical models. 
\ion{O}{6} is difficult to produce through 
photoionization, since it requires photons with energy greater than
114 eV, and is therefore most likely collisionally ionized.
It peaks in abundance at $3\times 10^5$ K, compared
to $2\times 10^5$ K, $1\times 10^5$ K, and $6\times 10^4$ K for
\ion{N}{5}, \ion{C}{4}, and \ion{Si}{4}, respectively.
\ion{O}{6} is either produced via cooling from higher temperatures or from heating. 
Several models have been proposed  to account for \ion{O}{6} in the ISM,
including shock heating, conductive heating, radiative cooling, and turbulent mixing
layers (TMLs) \cite[][and references therein for a summary]{shull79,borkowski90,slavin93,spitzer96}.
These models make different predictions for the various ratios of the highly ionized species, 
in particular, for convective heating $0.1 \la N($\ion{C}{4}$)/N($\ion{O}{6}$) \la 0.3$, 
for radiative cooling, $0.1 \la N($\ion{C}{4}$)/N($\ion{O}{6}$) \la 0.5$,
and for TMLs, $N($\ion{C}{4}$)/N($\ion{O}{6}$) \ga 1$ \citep[][and references therein]{spitzer96,sembach01}. 

\begin{figure*}[!th]
\begin{center}
\includegraphics[width=15.truecm]{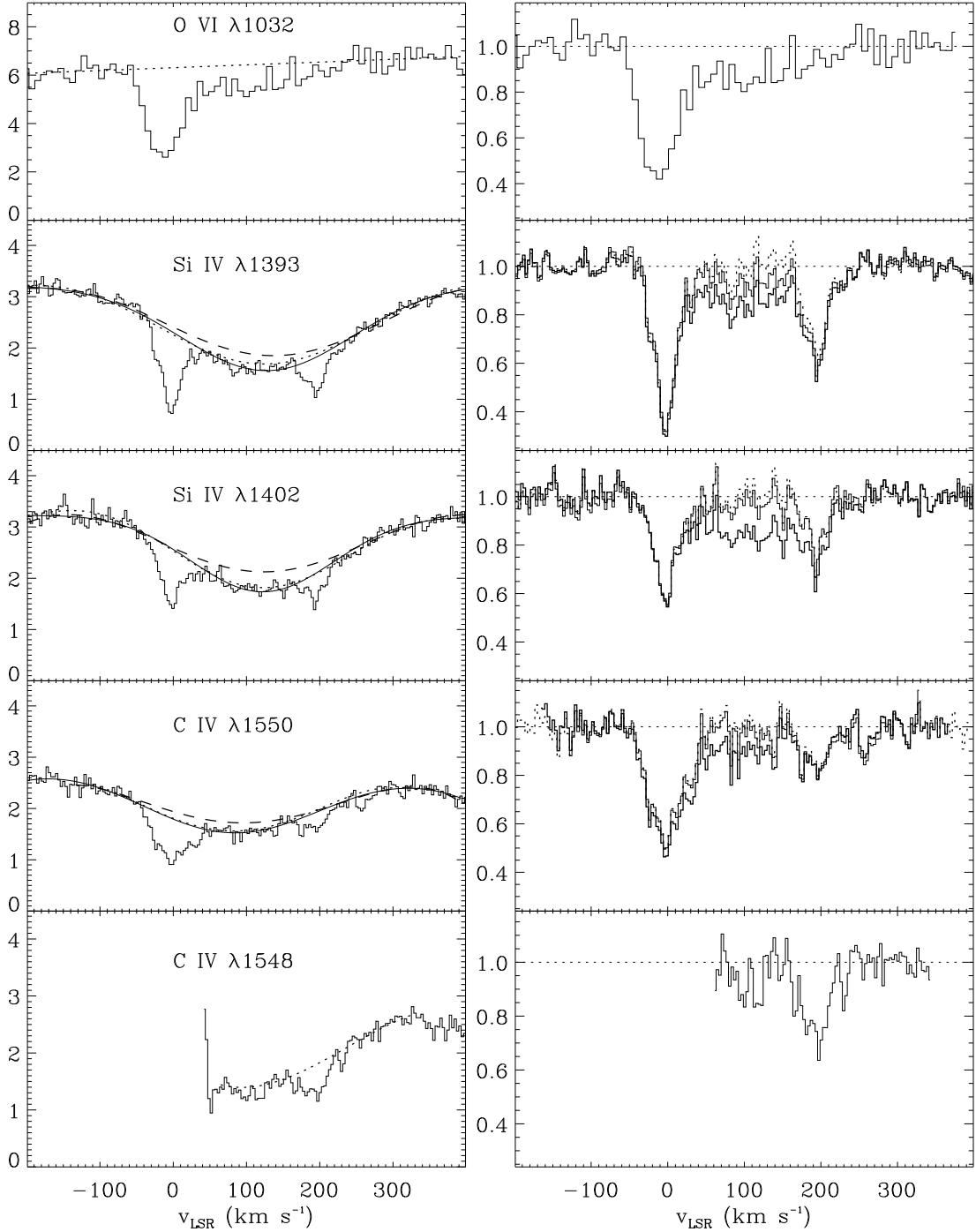}
\caption{Comparison of the highly ionized interstellar species toward DI\,1388.
{\em Left panel}: Profiles in flux unit ($10^{-13}$ erg\,cm$^{-2}$\,s$^{-1}$\,\AA$^{-1}$) versus the LSR velocity.
The continua are indicated by the solid line (gaussian fit), dotted and dashed line (polynomial fit).
{\em Right panel}: Normalized flux versus the LSR velocity. The dotted lines result from the gaussian fit continuum,
while the solid lines result from the polynomial fit continuum.
} 
\label{fig5a0}
\end{center}
\end{figure*}

As seen in Figure~\ref{fig1a}, the \ion{O}{6} absorption profile toward the DI\,1388 sight line is 
strong for the Galactic gas between $-50$ and 50 \km\, and becomes weaker
at positive velocity up to $\sim$250 \km. There is therefore
no clear separation between the Galactic disk/halo and MB components. 
We note also that some of the observed \ion{O}{6} absorption could have a stellar-wind 
origin \citep{lehner01c},  yet the interstellar \ion{O}{6} origin seems more plausible
as wind features typically have full-width larger than few hundreds \km.
Moreover, \ion{O}{6} absorption profiles toward the Magellanic Clouds are generally very broad
with the Galactic and LMC/SMC components being blended \citep{howk02,hoopes02}. 
In Figure~\ref{fig5a0}, we compare the absorption
profiles of the highly ionized species observed in the \fuse\/ 
(\ion{O}{6}) and STIS \citep[\ion{C}{4} and \ion{Si}{4}, see also][]{lehner01a} bandpass. 
The local component ($[-50,50]$ \km) is strong and well defined for all the highly 
ionized species. But at higher positive velocities, \ion{C}{4} and \ion{Si}{4} have 
a strong absorption MB component ($[155,245]$ \km) while \ion{O}{6} is relatively weak, 
whereas at velocities $\la 155$ \km, the \ion{O}{6} profile becomes deeper.
The difference in the profiles of  \ion{C}{4} and \ion{Si}{4} with 
respect to \ion{O}{6} at $\sim$200 \km\ could be due to that some of the 
\ion{C}{4} and \ion{Si}{4} originated from  photoionization. 

As \ion{O}{6}, \ion{C}{4}, and \ion{Si}{4} are usually related, 
in view of the broad extent of the \ion{O}{6} profile, we decided
to redo the analysis of \ion{C}{4} and \ion{Si}{4} by considering 
different possible continua. This is shown in Figure~\ref{fig5a0}, 
where the deep Gaussian and polynomial continua fit
a stellar line that goes to the bottom of the absorption profiles. Shallower stellar 
profiles also seem possible to allow
the interstellar \ion{C}{4} and \ion{Si}{4} to be present at all velocities
as observed for \ion{O}{6}. The resulting column density measurements are summarized
in Table~\ref{t1a} for different velocity intervals.\footnote{\citet{lehner01a} 
used only a Gaussian fitting method but they also allow possible renormalization of the profiles. 
For \ion{Si}{4} we find similar results, while \ion{C}{4} is slightly lower, though in 
agreement within the uncertainties. However, their main result was from the direct
comparison of apparent column density profiles of \ion{C}{4} and \ion{Si}{4}. Their derived
$N($\ion{C}{4}$)/N($\ion{Si}{4}) ratio gives a similar \ion{C}{4} column density,
assuming their \ion{Si}{4} column density.} The errors
presented in this Table were obtained following the method outlined by \citet{sembach92}.
The differences between the various measurements reflect systematic differences that
result from various choices for the wavelength intervals or methods defining
the (unknown) location of the true continuum. Note that the \ion{C}{4} $\lambda$1548
is at the edge of the \'echelle order and only the MB component could be (partially) retrieved;
column densities for the \ion{C}{4} doublet are in good agreement. The 
\ion{Si}{4} doublet gives very similar results. However, saturation effects 
are difficult to estimate due the uncertainty in the continuum placements.

In Table~\ref{t1b}, we present the adopted column densities and column density 
ratios of the highly ionized species. For \ion{C}{4} and \ion{Si}{4} and for 
all the column density ratios, the error bars indicate the range of possible values
rather than a 1$\sigma$ uncertainty, yet the upper error could be larger if the stellar line
is even shallower. We did not take into account the high continuum for \ion{C}{4} $\lambda$1550
in these adopted results, but this does not change the main conclusions (see below).

The TML model best reproduces the observed ratios of 
\ion{C}{4} to \ion{O}{6} and \ion{Si}{4} for the main MB component between
155 and 245 \km. If \ion{C}{4} and \ion{Si}{4} column densities
are underestimated, the agreement with the TML model would be strengthened, since it requires
$N($\ion{C}{4}$)/N($\ion{O}{6}$) \ga 1$.  The 
different gas phases observed at close or indistinguishable velocities 
also give some evidence for such a model because cold gas, partially ionized gas and hot gas 
phases are encountered along this sight line \citep[see Figure~1 in][]{slavin93}.
These models usually assume solar abundances or Galactic depletion and it would 
be interesting to investigate in more details the effect of low metallicity
as observed in the MB. We note, however, that neither the stellar abundances from  a 
few early-type stars \citep{rol99} nor the relative ISM abundances along this sight 
line show any anomalies in the relative abundance of the metals. For lower 
MB velocities (100--155 \km), the ratios are very uncertain due to the continuum placement,
but suggest a combination of TML and conductive interfaces. The HVC component is  
very uncertain (see \S~\ref{hvc}). The ratios of highly ionized species for the strong local component
is in agreement to what is found in the Galactic disk and halo; and
the radiatively cooling fountain provides a reasonable description of
the observed ratios toward DI\,1388 \citep[see,][and references therein]{sembach99,spitzer96}. 

The amount of \ion{H}{2} can be roughly estimated along this sight line from the \ion{O}{6} measurement
via $N($\ion{H}{2}$) = N($\ion{O}{6}$) ({\rm O/H})^{-1}_{\rm MW} Z^{-1} f^{-1}_{\rm O\,VI}$. 
Assuming an \ion{O}{6} ionization fraction $f_{\rm O\,VI} \le 0.2$ \citep{tripp00}, 
a present-day metallicity $Z = 0.08 Z_\odot$ \citep{rol99}, and a Galactic interstellar abundance
$ {\rm O/H} = (3.43 \pm 0.15)\times 10^{-4}$
(see \S~\ref{h2}), $\log N($\ion{H}{2}$) \approx 18.40 $ dex in the range
$[155,245]$ \km\ and $\log N($\ion{H}{2}$) \approx 18.85 $ dex in the range
$[100,245]$ \km, implying that neutral gas dominates along this sight line.

\subsection{Depletion of the gas}\label{dep}
\begin{figure*}[!th]
\begin{center}
\includegraphics[width=12.truecm]{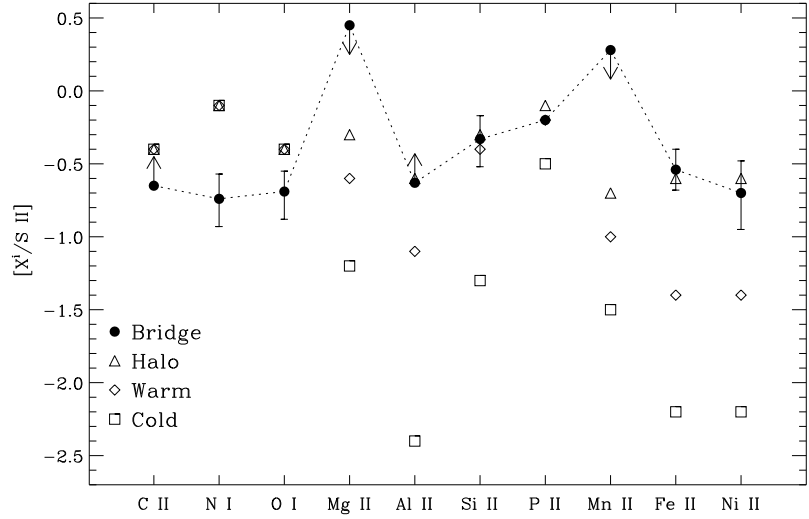}
\caption{Relative gas-phase abundances for the MB (DI\,1388 sight line), with respect to
sulfur (see Table~\ref{t3}), compared to relative abundances found in Galactic cold, warm and halo clouds.
Error bars are $1\sigma$. The upward (downward) arrows indicate lower (upper) limits.
The value for \ion{P}{2} remains uncertain.  The dotted line
represents the MB depletions corrected from the  underlying (undepleted) total elemental relative 
abundance patterns of the SMC with respect to the Galaxy.} 
\label{fig6}
\end{center}
\end{figure*}

In the Galactic interstellar gas, the underabundances of elements
along various sight lines are usually attributed to the depletion
of these elements into dust. The nucleosynthetic history of the 
gas can also play a role, especially in a low metallicity and tidally disrupted gas.
Yet, toward DI\,1388, \citet{lehner01a} showed that 
the principal factor for deficiency of heavy elements 
is certainly depletion onto dust grains. Here, we combine 
{\em FUSE}\/ and STIS results to develop more insight
into the depletion and possible ionization correction.
Since \ion{S}{2} is only modestly depleted in the 
Galactic ISM, we used it as the reference ion. 
We compared the elements using the logarithmic normalized gas-phase abundance,
\begin{equation}\label{dg}
\left[\frac{X^{i+}}{{\rm S}}\right] = \log\left(\frac{X^{i+}}{{\rm S^+}}\right) - \log\left(\frac{X}{{\rm S}}\right)_{\rm c} \,,
\end{equation}
where $X^i$ is the ion under consideration, and $(X/{\rm S})_{\rm c}$ is the ratio for cosmic 
abundances and $X/{\rm S}$ is used for $N(X)/N({\rm S})$. Equation~\ref{dg} 
assumes that the ion $X^{i+}$ is the dominant form of element $X$. Since 
$N($\ion{S}{2}$)/N($\ion{S}{3}$)\sim 4.6$, \ion{S}{2} is the dominant form 
of S.
Figure~\ref{fig6} (see also Table~\ref{t3}) shows 
the comparison in abundance pattern  between 
the different Galactic environments and the results for the MB
toward DI\,1388, with both the {\em FUSE}\/ and STIS data.
The \ion{O}{1} column density is now better constrained compared 
to the study of \citet{lehner01a}, and the differences
observed with the Galactic environment could be mainly due
to some ionization effect. 
Similarly, \ion{N}{1} is still very deficient; yet taking into
account the ionized component (\ion{N}{2} and \ion{N}{3}), the depletion 
of N ($>-0.4$ dex) is roughly similar to the Galactic environment. 
The tentative measurement of \ion{P}{2} (at worst it should be an upper limit) also suggests a halo-like
depletion for the MB. Therefore, it seems that the dominant factor 
describing the MB  gas-phase abundance pattern {\em toward} DI\,1388 appears to be 
primary depletion of the elements into dust and secondary ionization for neutral species (\ion{O}{1},
\ion{N}{1}, and \ion{Ar}{1}) rather than nucleosynthetic history,
and the depletion pattern follows the pattern seen in the Galactic halo.

\section{DGIK\,975 sight line and comparison with DI\,1388 sight line}\label{dgik975}
\subsection{Weakly ionized species}
The data toward DGIK\,975 are much less reliable,
yet several species were detected:
\ion{C}{2}, \ion{N}{1}, \ion{N}{2}, \ion{O}{1},
\ion{O}{6}, \ion{Si}{2}, \ion{Fe}{2}, and \ion{Fe}{3}. 
This list hints  that as for the DI\,1388 sight line, this sight line
has different gas phases, with species dominant 
in neutral gas (\ion{N}{1}, \ion{O}{1}), some 
signatures of weakly ionized gas (\ion{N}{2}, \ion{Fe}{3}) 
or highly ionized gas (\ion{O}{6}). Comparisons of the
profile in Figure~\ref{fig2} suggest that, as toward DI\,1388,
several clouds are present along the sight line because
\ion{N}{1} is mainly observed at $\sim$170 \km, while for the
other species, at lower velocities, one or several clouds
are suggested.

The \ion{Fe}{2} column densities are consistent with each other within the 
errors. However, the data have low S/N levels, a resolution of 
$\sim$20 \km\ and the lines are very strong, which might suggest
some saturation effect.

\subsection{Highly ionized species}
\begin{figure*}[!th]
\begin{center}
\includegraphics[width=15.truecm]{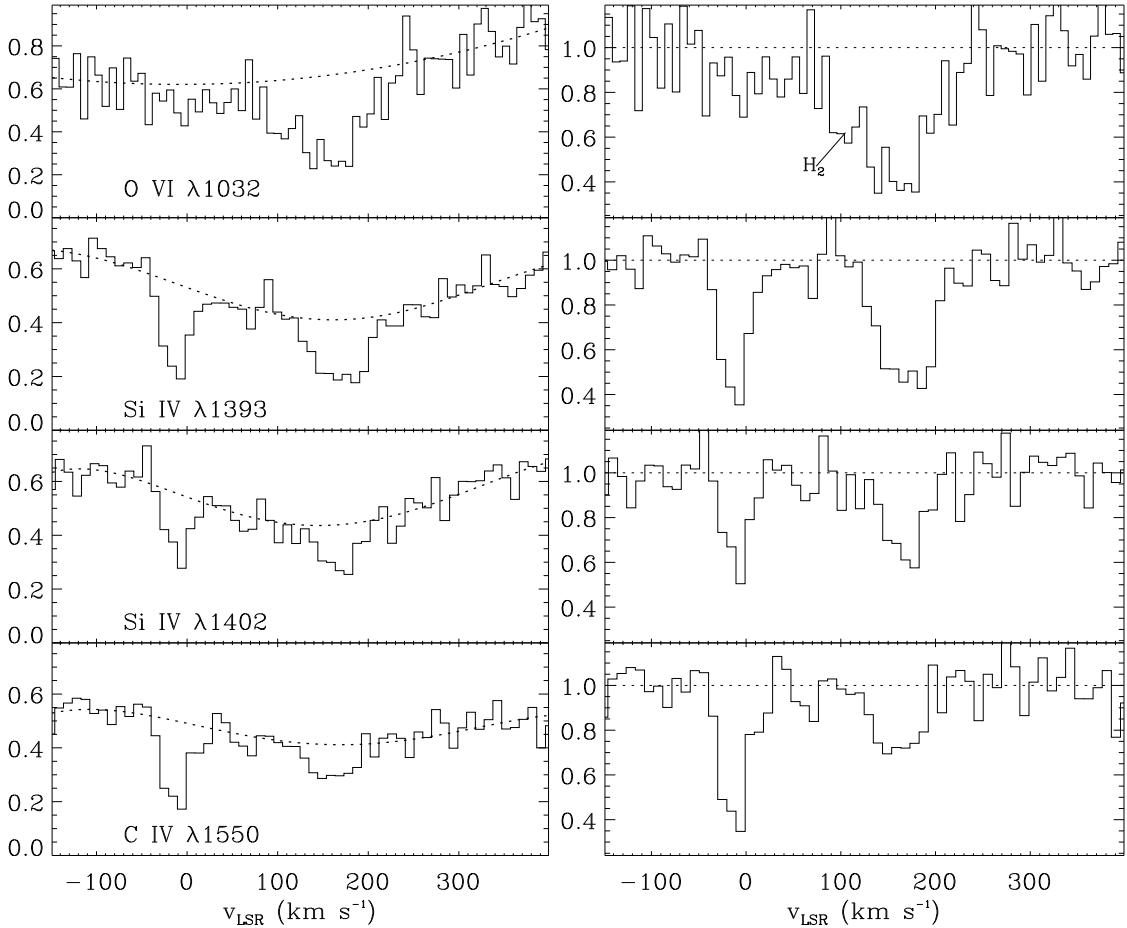}
\caption{Comparison of the highly ionized interstellar species toward DGIK\,975.
{\em Left panel}: Profiles in flux unit ($10^{-13}$ erg\,cm$^{-2}$\,s$^{-1}$\,\AA$^{-1}$) versus the LSR velocity.
The continua are indicated by the dotted line (polynomial fit).
{\em Right panel}: Resulting normalized flux versus the LSR velocity. A possible detection
of an highly ionized cloud is observed at $\sim +70$ \km.
} 
\label{fig5b}
\end{center}
\end{figure*}

{\em HST}/STIS observations were obtained toward DGIK\,975 
(in the same program as the observations of DI\,1388) but the data were of 
too low S/N ratio ($\la 5$ for most of the \'echelle orders)
for a detailed analysis \citep{lehner00}. However, since strong \ion{O}{6}
absorption is observed, it is 
interesting to try to estimate the amount of
\ion{C}{4} and \ion{Si}{4} along this sight line.
\ion{C}{4} and \ion{Si}{4}  are much broader than the 
neutral or weakly ionized species, and we can therefore
rebin the STIS E140M data by 3 to about the {\fuse}\/ spectral resolution
in order to improve the S/N without losing too much information. 
Once rebinned, the {\em HST}\/ data typically have a S/N
of about 10--12. In Figure~\ref{fig5b}, we present the different 
absorption profiles for \ion{O}{6}, \ion{C}{4}, and \ion{Si}{4} with
the chosen continua. Result of measurements is presented in Table~\ref{t2a}.
We assumed that there is no contribution from the stellar photosphere.
However, the projected rotational velocity
($v\sin i \sim 80$ \km) is relatively low and the radial 
velocity is $v_{\rm LSR} \sim 210$ \km\ so that some of the stellar
and interstellar lines could be blended. In this case, the $N($\ion{C}{4}$)/N($\ion{O}{6})	
and $N($\ion{Si}{4}$)/N($\ion{O}{6}) ratio presented in Table~\ref{t2a} should
be considered as an upper limit.

In Figure~\ref{fig5a}, the \ion{O}{6} apparent column density profiles for DI\,1388
and DGIK\,975 are compared. Toward
DI\,1388 there is no clear separation between MB and Galactic disk-halo
gas. The dashed line in Figure~\ref{fig5a}
shows a  Gaussian fit to the profile of \ion{O}{6}
toward DGIK\,975 with FWHM of $83 \pm 15$ \km, implying 
a temperature of about $2 \times 10^6$ K if the width of the line
is due to thermal broadening alone. For \ion{C}{4} and \ion{Si}{4}, the 
Gaussian FWHMs are $\sim 52$ and $35$ \km, respectively, corresponding
to a temperature of $\sim$$7 \times 10^5$ K. In collisional ionization 
equilibrium the \ion{O}{6}, \ion{C}{4}, and \ion{Si}{4} peak in abundance at lower temperatures 
(see \S~\ref{hot}),
so if most of the gas is at this temperature, either large turbulent
(non-thermal) motions or multiple components 
are required to explain these line broadenings. Because 
the column density is weaker toward DI\,1388, we can not 
fit reliably the \ion{O}{6} line, but its large extent also suggests 
a large temperature. For \ion{C}{4} and \ion{Si}{4}, the continuum
placement is uncertain but suggest a temperature $\ga 2 \times 10^5$ K. 
Both the LMC and SMC surveys of \ion{O}{6} absorption showed 
similar broad lines  \citep{howk02,hoopes02}.

While the amount \ion{O}{6} can not be directly compared 
in the MB because the location in depth of the stars is
unknown, $N($\ion{C}{4}$)/N($\ion{Si}{4}) is similar for both sight lines,
but $N($\ion{C}{4}$)/N($\ion{O}{6}) (and $N($\ion{Si}{4}$)/N($\ion{O}{6}))
is disparate.
This holds even if there is stellar contamination,
since toward DI\,1388 the  $N($\ion{C}{4}$)/N($\ion{O}{6}) and $N($\ion{Si}{4}$)/N($\ion{O}{6})
ratios would be lower limits, while toward DGIK\,975, they would be
upper limits. This suggests variation of the \ion{O}{6} column density
across the MB. In contrast, even if the Galactic 
component of hot gas varies by a factor 2.4  
between these sightlines \citep[confirming the recent results of][]{howk02a},
the ratio of the highly ionized
species agrees roughly within the uncertainties, suggesting that
similar mechanisms for the production of highly ionized ions occur
toward both sight lines in the Galactic gas.

Comparing with the theoretical models (see \S~\ref{hot}) toward DGIK\,975, 
a radiatively cooling flow or a combination in the right proportion
of TMLs and conductive interface describes
the derived ratios reasonably well for the MB component.

\section{The High-Velocity Cloud HVC $291.5-41.2+80$}\label{hvc}
Using the ultraviolet STIS spectrum of DI\,1388,
\citet{lehner01b} detected an HVC at $v_{\rm LSR}$ $\sim +80$ \km, 
as well as  two very low-column HVCs at $\sim +113$ and +130 \km.
The \ion{O}{1} and \ion{N}{1} deficiencies with respect to \ion{C}{2}
and the presence of ionized elements (\ion{S}{3} and \ion{Si}{3}) 
indicated that HVC\,$291.5-41.2+80$  consists of a warm, mainly ionized 
medium at high Galactic latitude.

\begin{figure*}[!th]
\begin{center}
\includegraphics[width=9truecm]{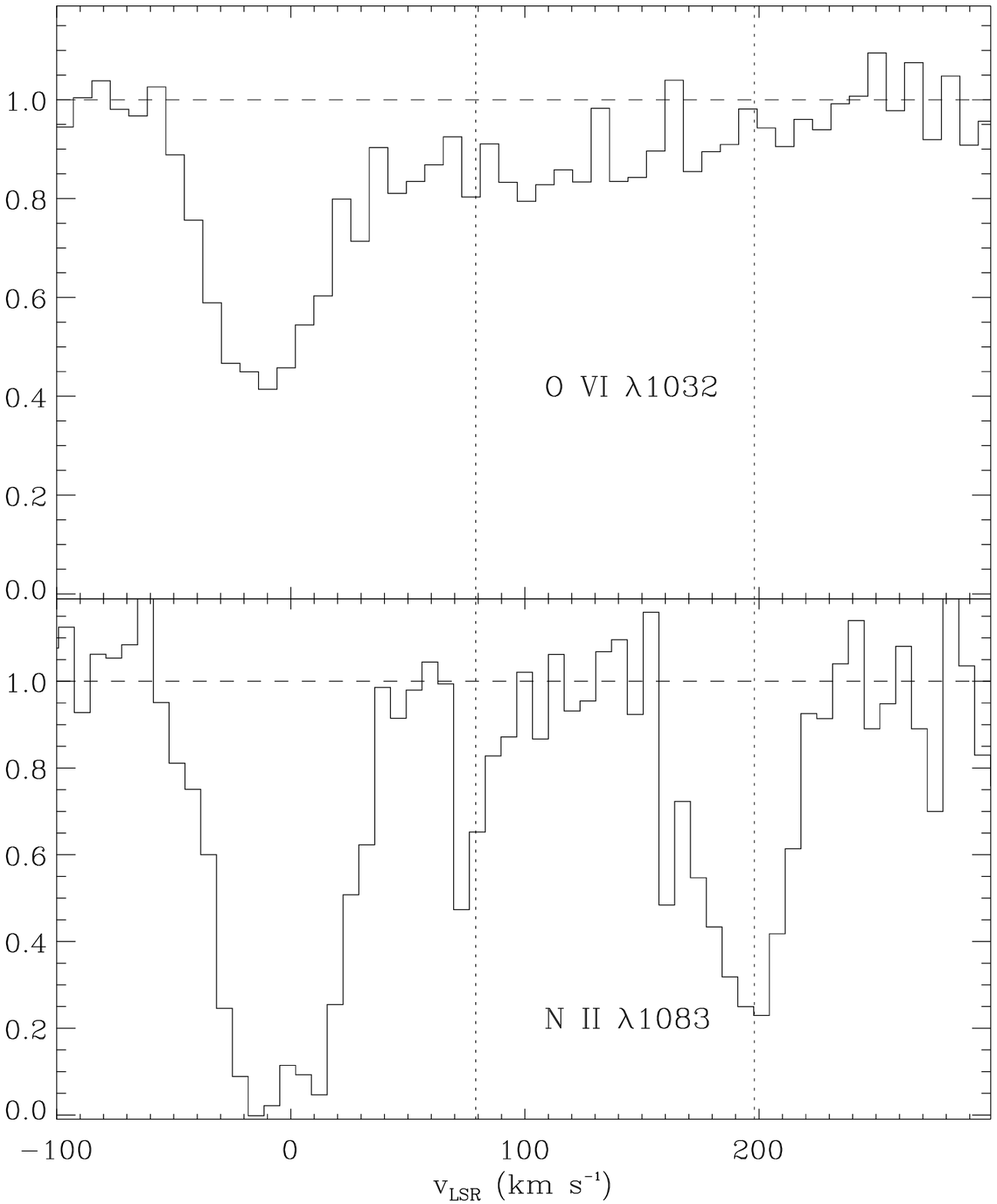}
\caption{Normalized profiles of absorption lines of \ion{O}{6} and \ion{N}{2}
in the \fuse\/ spectrum of DI\,1388. 
Absorption from the Magellanic Bridge is detected at $\sim 200$ \km, while 
the HVC is at $\sim 80$ \km.} 
\label{fig1a}
\end{center}
\end{figure*}

\begin{figure*}[!th]
\begin{center}
\includegraphics[width=9.truecm]{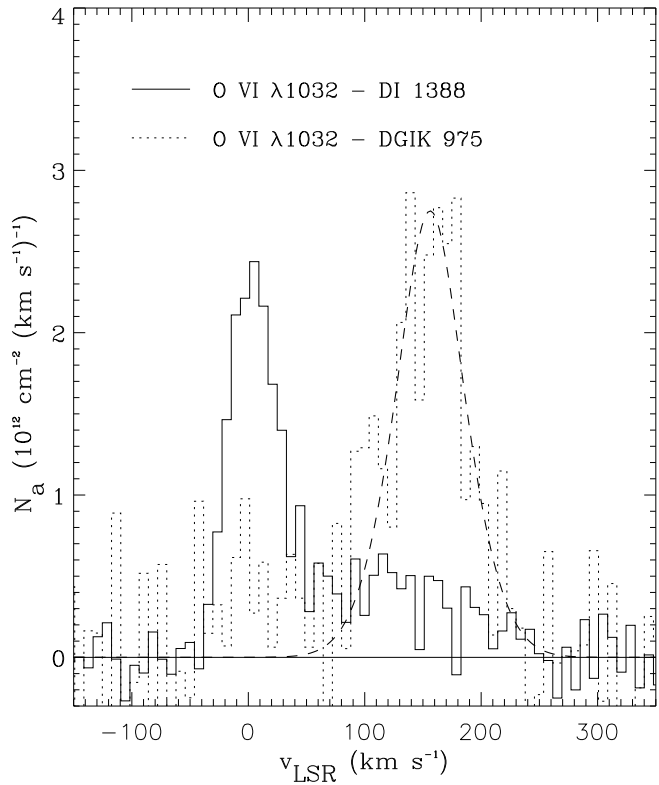}
\caption{Comparison of the apparent column density profiles ({\em histograms}) for \ion{O}{6} in the direction 
of DI\,1388 and DGIK\,975. The dashed line is a Gaussian fit to the \ion{O}{6} absorption line.} 
\label{fig5a}
\end{center}
\end{figure*}

In the \fuse\/ bandpass, this HVC is detected in absorption via \ion{N}{2}
and \ion{O}{6}, and their normalized profiles are shown in 
Figure~\ref{fig1a}. No \ion{Fe}{3} is detected but this is not 
surprising as the \ion{Fe}{3} stellar line is strong and the interstellar HVC component
is lost in it. The apparent column density of \ion{N}{2} is $13.44 \pm 0.12$ dex,
confirming that this HVC is largely ionized \citep[$\log N($\ion{N}{1}$)< 12.4$ dex,][]{lehner01b}.
No \ion{N}{3} is detected, and we estimated a 3$\sigma$
upper limit of $< 13.5$ dex. 

As discussed in \S~\ref{hot}, there is no clear 
separation between the different components in the 
\ion{O}{6} absorption profile (see Figures~\ref{fig1a} and \ref{fig5a}).
We used \ion{N}{2} to determine the velocity range (70--100 \km) over
which the \ion{O}{6} intensity profile is integrated. 
In Table~\ref{t1b}, we summarize this measurement along with the results
on \ion{C}{4} and \ion{Si}{4}. The latter remains uncertain due 
to continuum placement (see \S~\ref{hot}). The \ion{C}{4}/\ion{O}{6} 
ratio is in agreement with values found in the Galactic disk and near
the low end value for the Galactic halo \citep{spitzer96}. \citet{zsargo02}
presented a more recent survey in the low Galactic halo, where they found
$N($\ion{C}{4}$)/N($\ion{O}{6})$< 1$
for most sight lines with Galactic height $z\la 4$ kpc.

The high ion ratios are consistent with the expectations of a radiatively cooling 
gas in a fountain flow  \citep{spitzer96,sembach99}. In a
such model hot gas is produced  by the cumulative effects 
of supernovae in the Galactic disk. It provides an explanation for the origin of HVCs 
in the low Galactic halo with positive velocity \citep[][and references therein]{wakker97}, 
such as HVC $291.5-41.2+80$. The location of the HVC in the lower part of the halo also provides 
an explanation of why the gas is mainly ionized, since a sufficient number of 
photons could leak out of the Galactic disk to cause its ionization.
Alternative scenarios are possible but would require the combination
of conductive interfaces and turbulent mixing layers.

We note also a tentative detection of an HVC at $\sim +70$ \km\ toward 
DGIK\,975 in the high ions (see Fig.~\ref{fig5b}). The S/N levels 
are too low to make any measurements  but would confirm that such
highly ionized clouds are common. 

\section{Discussion}
\subsection{Molecules and star formation in the Magellanic Bridge}\label{disc1}
\citet{demers98} found using deep $BV$ CCD photometry that the stars in form of stellar 
clusters in the MB were formed some 10 to 25 Myr ago. Spectroscopic
studies of some of these stars confirm that they are massive young
stars but with a metallicity $Z \approx 0.08 Z_\odot$ \citep{rol99},
a factor $\sim$3 lower than the present day metallicity of the SMC. 
Their location in the MB also rejects the idea that these stars
could be runaway stars from the SMC since they could not travel such large
distances over their short life-times. It is generally accepted
that the MB was most likely pulled from
the wing of the SMC some 200 Myr ago,
during a close encounter between the two Clouds, whereas the Magellanic 
Stream was created by a SMC-LMC-Galaxy close encounter $\sim1.5$ Gyr ago
\citep{gardiner}.
Searches for stars in the Stream have, however, produced largely negative 
results \citep[see][and references therein]{irwin90}. 
Interstellar study toward two sight lines in the Stream
indicates an abundance similar to the
abundance in the SMC, $Z \approx 0.25 Z_\odot$
\citep{lu98,sembach01,gibson00}.
 Yet, because the present-day MB abundance
is a factor $\sim$3 lower than the present day metallicity of the SMC,
\citet{rol99} postulated that the MB was formed either much earlier ($\sim$8.5 Gyr ago)
from more primordial gas, or  some 0.2 Gyr ago from a mixture of SMC gas and an 
unenriched (primordial) component. Toward both the Magellanic Stream and Bridge,
molecules were found, yet because star formation occurs in the MB and not in 
the Stream, the physical conditions must be different. Even though young stellar 
clusters exist in the MB \citep{demers98}, their formation is not very
efficient and only happens in recent times, since the MB gas is poor in
metals. \citet{christ97} suggested that the MB was formed earlier
than the Stream, in order to explain why star formation occurs only in the MB. 
In such a scenario, the low metallicity is explained 
by the tidal stripping of gas from SMC in its earlier chemical 
evolution, and stars form late, only after tidally induced cloud mergers
have had a chance to occur in substantial numbers, and to produce 
some massive clouds. Such scenario is supported by
some recent \ion{H}{1} Parkes All Sky Survey (HIPASS) data of the entire 
Magellanic Stream which indicate dual filaments that are likely to be 
relics from gas stripped from the SMC and the MB \citep{putman02}. 
These observations imply that at least some part 
of the Stream could be younger than the MB, and that the MB
could be older than usually assumed  \citep[see,][]{putman02}. 
Pairing of galaxies are very common \citep{barnes92}, 
and the SMC and LMC could have been paired since the beginning of 
their existence, and therefore some part of the MB, if not all, 
could be much older than previously thought.

The discovery of molecular hydrogen in the MB implies that
it is formed {\em in situ} and/or was tidally stripped 
from the SMC.
Formation of H$_2$ is thought to occur on interstellar
dust grains when hydrogen atoms are adsorbed onto the grain 
surface and react. An hydrogen molecule is then ejected in the gas
phase \citep[e.g.,][]{hollenbach71}. Therefore the formation 
mechanism is related to the quantity, type of grains,
gas density, and metallicity. Dissocation of  H$_2$ depends on
the intensity of the ultraviolet radiation field, the presence 
of hot gas, and the possible effect of shocks. 
Grains serve as catalysts for the formation of H$_2$ but also as 
shields against the ultraviolet radiation. \citet{tumlinson02} found that
for the LMC and SMC low metallicity gas differences 
in the formation and destruction balance of H$_2$: grain formation 
rates about 1/3 to 1/10 of the Galactic value and ultraviolet radiation 
field 10 to 100 times the Galactic mean value. The 
H$_2$ formation can be written as $t = 1/(Rn_{\rm H})$ \citep[e.g.,][]{shull82}, where 
$R$ is the formation rate coefficient, ($R= 1-3\times 10^{-17}$ cm$^{3}$\,s$^{-1}$
for Galactic conditions, $R \approx 0.3\times 10^{-17}$ cm$^{3}$\,s$^{-1}$ for LMC, SMC
conditions) and $n_{\rm H}$ the hydrogen column density. Assuming LMC, SMC 
conditions for the MB, $t = 11/n_{\rm H} $ Gyr. Along the DI\,1388 sight line, 
$n_{\rm H}$ cannot be high enough to lower $t$ to the believed 200 Myr year old of the MB, 
and therefore from those assumptions most of the observed  H$_2$ has survived the tidal stripping;
unless if the MB is much older.

Alternatively, H$_2$ could have  (partially)
formed in small compact and dense cloudlets within the MB. 
Formation of clusters of massive stars occurs in the MB 
\citep[][and references therein]{rol99,demers98}, and if the 
formation processes are similar to that observed in the Galaxy, this requires
very high density clouds and cloud-cloud collisions \citep{scoville86,evans99}. 
In this picture H$_2$ can form rapidly in dense regions and disperse in 
more diffuse clouds. Therefore the observed diffuse molecular clouds
along the DI\,1388 sight line could trace or be the remnant of denser
molecular clouds, the latter being the place where star formation 
occurred. The detection of cold \ion{H}{1} cloud with spin temperatures
between 20 and 50 K in the MB by \citet{kob99} also supports the idea
that some regions must be at high densities ($\ga 100$ cm$^{-3}$).
The pressure in the MB must be low on average, yet tidal interactions between 
the SMC and LMC could have created strong enough perturbations in the overall 
low density MB gas to produce high density peaks, conditions not existing 
in the Stream because possibly the interaction in the MB involves closer galaxies 
and/or is over a longer time.

\subsection{N/O ratio in the Magellanic Bridge}\label{disc2}
The nucleosynthetic origin of N has been subject 
of large debate, but its understanding is fundamental
for comprehending the chemical evolution of galaxies
\citep{vila93,henry99}. Nitrogen is mainly produced in
the six steps of the CN branch of the CNO bi-cycle
within H burning stellar zones, where $^{12}$C serves
as a catalyst. First generation stars produce their own
C during the He burning phase, 
and N production must be fairly independent 
of the initial composition of the star of which
it is synthesized, and the synthesis is said {\em primary}.
Beyond the first generation of stars, the gas from which these  stars
formed is already polluted with C and O. The amount of N 
in this material will be proportional of its C abundance, and 
the N synthesis is said to be {\em secondary} in this case. 
Thus, primary nitrogen is independent of the metalliticity,
and secondary nitrogen increases with increasing metallicity.
Most measurements of N/O in our Galaxy and nearby Galaxies 
are made in stars and \ion{H}{2} regions
and only few direct measurements in the gas are available in absorption, 
and therefore our measurement in the MB is particularly notable.

Toward DI\,1388, the MB N/O ratio is $[{\rm N/O}] = -0.05 \pm 0.02$ (from 
\ion{N}{1} and \ion{O}{1} column densities), i.e. near a solar
value. \citet{lehner01a} suggested that N might be deficient in the MB,
however, they also noted that this could be mainly due to photoionization effect. 
The {\fuse}\/ data allow us to have precise measurements of \ion{N}{1} and \ion{O}{1} 
but also give access to ionized N via \ion{N}{2} and \ion{N}{3}. If \ion{N}{1}
is preferentially ionized with respect to \ion{O}{1},
$[{\rm N/O}]$ would be solar or even above solar. 
The N/O ratio in the gas is in agreement with the results on the photospheric abundances 
of the B-type stars analysis where $[{\rm N/O}] \approx -0.08$ 
\citep[from a differential abundance analysis,][]{rol99}. 
In contrast, in both the LMC and SMC, N in the \ion{H}{2} regions is 
systematically underabundant by a factor 4--6 compared to the main sequence
B-type stars \citep[see,][]{garnett}. 
An analysis of the \fuse\/ data using
absorption lines could help to constrain N/O in the ISM of the LMC and SMC,
and therefore help to comprehend these differences. 

\citet{henry00} compiled a survey of Galactic and 
extragalactic \ion{H}{2} regions and stars to analyse the behavior of N/O
as a function of O/H. They observe a plateau (though with some scatter) 
of N/O measurements when $[{\rm O/H}] < -1$ suggesting independence 
of N/O with metallicity, consistent with primary nitrogen formation; 
and when $[{\rm O/H}] \ga -1$,  a rise with a large scatter of N/O with 
metallicity is observed, more suggestive of secondary nitrogen formation.
The MB metallicity \citep[$-1.1 \pm 0.1$ dex for C, N, O, Mg, and Si in the photosphere 
of main sequence B-type stars,][]{rol99} is at this apparent boundary. 
\citet{henry00} found, at the MB metallicity, N/O lower by a factor 3 to 10 compared
to the MB. Yet, the typical error bars for these measurements are 
large, typically $\sim \pm 0.4$ dex for $\log ({\rm N/O})$ and $\sim \pm 0.2$
dex for $\log ({\rm O/H})$, and within the error bars few measurements
of N/O in \ion{H}{2} regions could have a solar value for a MB metallicity.
In the damped Ly$\alpha$ systems, at  MB metallicity (or slighly higher or much lower
than the MB metallicity), N/O is always at least a factor 3 lower than solar 
\citep{prochaska02,lu98b}. 

From both the studies of the gas and the massive stars,
N/O is much higher in the MB than N/O values in the damped Ly$\alpha$ systems, implying that 
N in the MB is more secondary than primary. Most of nitrogen is formed
in intermediate mass-stars (1--8 M$_\odot$) with roughly a charateristic 
lag time of 250 Myr \citep{henry00}. If the MB is 200 Myr old (see \S~\ref{disc1}), 
most of the nitrogen must have come from an N-enriched region of the SMC. \citet{henry00}
argue that since the distribution of N/O at a single O/H value appears to be clustered
toward low N/O values, the high values of N/O might be due to nitrogen enrichment
by Wolf-Rayet stars or luminous blue variables.

\section{Summary}
We have presented \fuse\/ observations of two early-type stars (DI\,1388 and DGIK\,975) in the 
low density and low metallicity gas of the MB. DI\,1388 is situated near the SMC, while 
DGIK\,975 lies near the LMC.  Galactic ($v_{\rm LSR} \sim 0$ \km), HVC ($v_{\rm LSR} \sim 80$ \km), 
and MB ($v_{\rm LSR} \sim 150$--200 \km) components are observed in absorption. 
The \fuse\/  spectra reveal that the MB gas shows a complex arrangement of
gas in different phases, including the
detection of molecules, neutral, weakly and highly ionized
species as observed in the interstellar medium 
of the Galaxy. 

Toward DI\,1388, the \fuse\/ observations provide
detections of molecular hydrogen and \ion{O}{6}. Numerous
transitions of the same species available in the \fuse\/
bandpass allow us to derive precise column density for 
\ion{O}{1}, \ion{N}{1}, and \ion{Fe}{2}. It gives also 
useful information on the amount of \ion{N}{2}, \ion{N}{3}, 
\ion{Ar}{1}, \ion{P}{2}, \ion{S}{3}, and \ion{Fe}{3}. Combining the previous 
STIS results \citep{lehner01a} with the present results
show that the relative abundance pattern with respect to S in the MB along the
DI\,1388 sight line can be attributed to varying degrees of depletion 
onto dust similar to that in halo clouds, despite an overall 
metallicity much lower, and 
the depletion pattern follows the pattern observed in the Galactic halo. 
For \ion{N}{1} (and \ion{Ar}{1}) the deficiency can be
explained by ionization effects, and the amount of \ion{N}{2} is at least as
large as the amount of \ion{N}{1}. 

The N/O ratio in the MB toward DI\,1388 is near solar, in agreement with 
the abundance exhibited by early-type stars \cite[see,][]{rol99}. This is a factor 3 or higher
than the values of N/O
in the damped Ly$\alpha$ systems, implying subsequent stellar 
processing (possibly from Wolf-Rayet stars or luminous blue variables) 
to explain this high N/O ratio for an overall low metallicity. 

The diffuse molecular cloud has a low column density ($\log N ({\rm H}_2) \approx 15.43$ dex),
yet two excitation temperatures are needed to fit the distribution of the different
rotational levels. We show that this is not uncommon in the Clouds (see Appendix). 
The low $b$-value for H$_2$ is similar to the value for \ion{N}{1}, whereas 
\ion{O}{1} and other species have larger $b$-values, implying that several clouds
are present along the DI\,1388 sight line, including, cold and warm neutral clouds, a
partially weakly ionized cloud, and highly ionized cloud.

Toward DGIK\,975, the situation is less clear due to lower quality data and
stronger lines (more subject to saturation effect). Yet the presence
of neutral, weakly and highly ionized species suggest that this sight line 
has also several gas phases as observed toward DI\,1388 or the 
interstellar gas in the Galaxy.

For the highly ionized species along the sight line to DGIK\,975, for the MB component,
very broad features of \ion{O}{6}, \ion{C}{4}, and \ion{Si}{4}
are observed, indicating  that either non-thermal motions are prevalent, or that multiple components 
of hot gas at different velocities are present. Toward DI\,1388, 
the profile in \ion{O}{6} is extended, but where the \ion{O}{6} absorption is the 
weakest (main component of the neutral gas at 198 \km)  \ion{C}{4} and \ion{Si}{4}
are the strongest (though some uncertainties remain due to continuum placement),
which indicates that \ion{C}{4} and \ion{Si}{4} could have partially originated
from photoionization. 
Similar large broadening were found in LMC and SMC \citep{howk02,hoopes02}.
Toward DI\,1388, a high ratio of \ion{C}{4}/\ion{O}{6} ($>1$) is found for the main 
component at 198 \km\ that could have originated  in a turbulent mixing layer.
Toward the SMC star Sk\,108, a similar high ratio was found \citep{mallouris01}. When considering 
the more extended absorption profiles toward DI\,1388 (in the range $[100,245]$ \km)
the \ion{C}{4}/\ion{O}{6} ratio is lower and more similar to the ratio
toward DGIK\,975. \ion{C}{4}/\ion{O}{6} varies 
within the MB but the \ion{C}{4}/\ion{Si}{4} is relatively constant (in comparison) 
for both sight lines.  Several sources (a combination of turbulent mixing layer, 
conductive heating, and cooling flow) may cause the hot ionized gas in the MB.

H$_2$ is observed in both the Magellanic Stream and the MB, yet the
formation of massive stars occurs only in the MB, implying significantly different physical
conditions between them. In the MB, some of the H$_2$ could have been
pulled out from the SMC via tidal interactions, but some also could
have formed {\em in situ} in dense clouds where formation of molecules occurs
faster. These dense clouds are possible sites of star formation.

Finally, this study has confirmed the results of \citet{lehner01b}
that the HVC at $\sim80$ \km\ is mainly ionized, but high ions
are also associated with it. The high ion ratios are consistent with a 
radiatively cooling gas in a fountain flow model. Such a
model provides an explanation for the origin of HVCs 
in the low Galactic halo with positive velocity. 
Alternative scenarios would require a combination
of conductive interfaces and turbulent mixing layers.

\acknowledgements
I thank the FUSE PI team for making these observations 
possible. I am grateful to Alex Fullerton for a careful reading
of the manuscript and for reprocessing the \fuse\/ data.
I thank Chris Howk for insightful suggestions and discussions.
The HST/STIS and AAT observations
were obtained by Francis P. Keenan and collaborators 
at Queen's University of Belfast.
This work is based on data obtained
for the Guaranteed Time Team by the NASA-CNES-CSA FUSE mission operated 
by the Johns Hopkins University. Financial support to U. S.
participants has been provided by NASA contract NAS5-32985.
This work is partly based on observations made with the NASA/ESA 
Hubble Space Telescope, obtained from the Data
Archive at the Space Telescope Science Institute, 
which is operated by the Association of Universities for
Research in Astronomy, Inc., under NASA contract NAS 5-26555. 
This research has made use of the NASA
Astrophysics Data System Abstract Service.

\appendix

\citet{spitzer74} found that for relatively low  H$_2$
column density in the Galaxy,  $N(J=0) \la 10^{15}$ cm$^{-2}$, a single
straight line with a slope characterized by $T_{\rm ex}$
generally fits all the observed levels $J=0,...,3$ or $J=0,...,4$.
At slightly higher column densities, $10^{15} \la N(J=0) \la 10^{15.5}$ cm$^{-2}$, 
two excitation temperatures are needed. This was also confirmed
by \citet{jenkins00a} toward $\epsilon$\,Ori. In their survey,
\citet{spitzer74} have 10 stars with $N(J=0) < 10^{15}$ cm$^{-2}$
and two with $10^{15} < N(J=0) \la 10^{15.5}$ cm$^{-2}$.
At higher column densities, two temperatures are 
generally needed to describe the populations in the
different $J$ levels. This is understood as a 
consequence of the density being high enough 
to ensure that collisions dominate over radiative
processes for low $J$ ($\la 1$) levels, which couples the levels
with the kinetic temperature. For higher $J$ ($\ga 4$), 
fluorescent pumping dominates and yields a higher excitation
temperature, whereas intermediate rotational levels ($J=2,3$)
are fixed by the rate of ultraviolet-pumping or can also be deexcited by
collisions due to their longer radiative lifetimes \citep{shull82}.

\begin{figure*}[!th]
\begin{center}
\includegraphics[width=16 truecm]{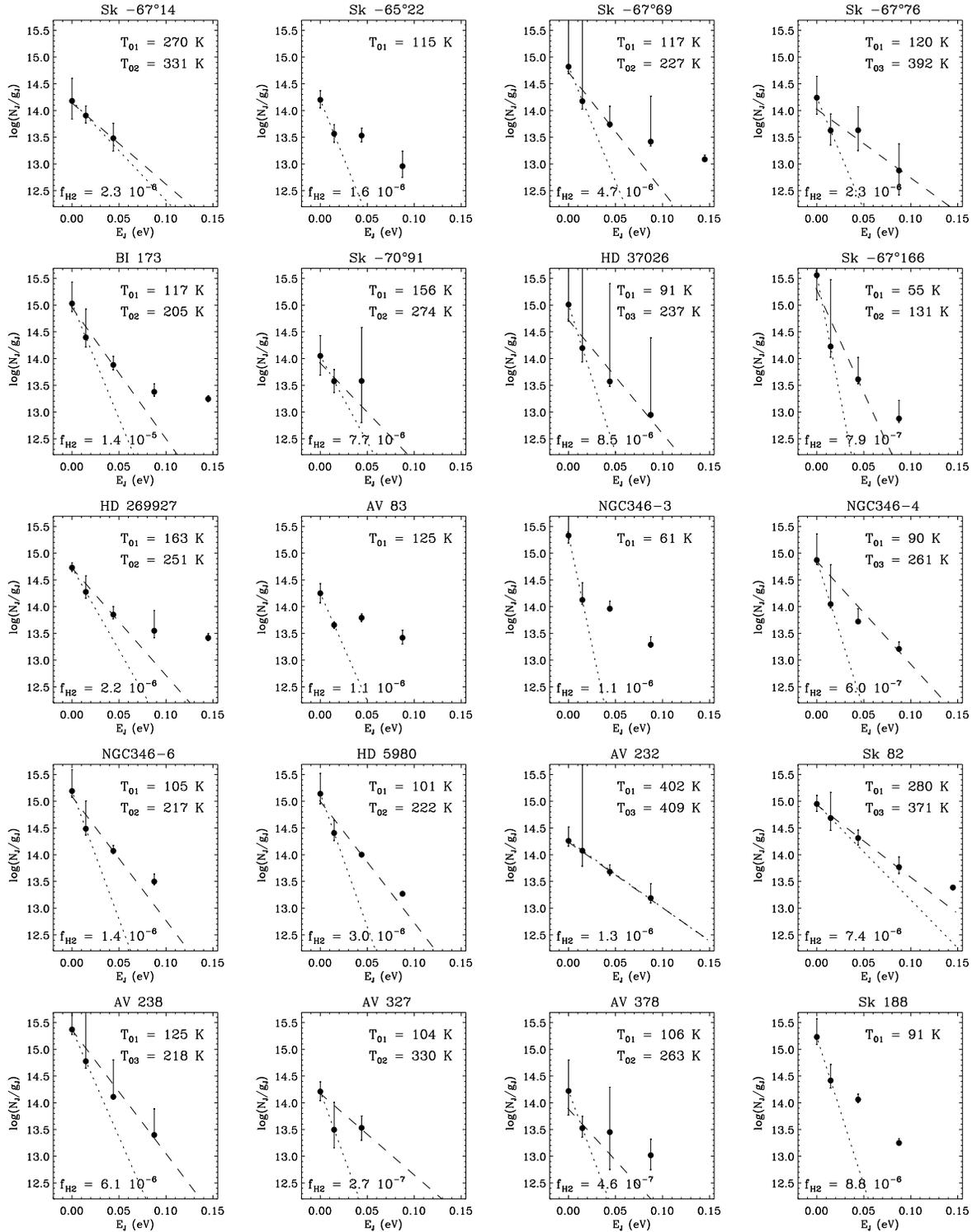}
\caption{H$_2$ column density ($N_J$) divided by statistical
weight ($g_J$) as a function of the excitation energy ($E_J$) for different rotational
levels $J$ in the LMC (first 9 panels) and the SMC. The $N(J)$ are from \citet{tumlinson02}. 
The dotted line represents the excitation temperature $T_{01}$ derived from 
Equation~\ref{eq1}. When present, the dashed line is a least square fit to the 
$N(J)$ and $T_{0j}$ (with $j>1$), showing the resulting excitation temperature and 
$j$, the number of levels used (see Appendix for more details). In the left corner 
of each panel, the molecular fraction is indicated.} 
\label{fig7}
\end{center}
\end{figure*}

Recently, \citet{tumlinson02} used {\fuse}\/ to measure
the H$_2$ column densities for different levels $J$ in the SMC and LMC. 
For comparison
with the Magellanic Bridge and the Galaxy, we reproduced here
their measurements for $N(J=0) \la 10^{15.5}$. 
We summarize the results in the form of an excitation diagram
in Figure~\ref{fig7} for 9 and 11 sight lines in the LMC
and SMC, respectively. In this Figure, the observed $N(J)$
are plotted with their 1$\sigma$ error bars estimated by
\citet{tumlinson02}. We indicate the excitation temperature
of the first two levels, $T_{01}$, derived directly from 
Equation~\ref{eq1}, and we plot the resulting {\em dotted } line
with a slope characterized by $T_{01}$. The data were then
fitted using a least-squares method by minimizing the $\chi^2$ error. 
The number of levels $j+1$ fitted simultaneously is indicated in Figure~\ref{fig7} 
by $T_{0j}$  ($  j>1 $).
When the goodness of fit was reasonable, the {\em dashed} line 
in Figure~\ref{fig7} shows the result from this least-squares fit;
when no {\em dashed} line is indicated, a proper straight line
could not be fitted within the estimated error bars. For reference we have
also indicated in this Figure the molecular fraction $f_{\rm H_2}$,
where both the \ion{H}{1} and total H$_2$ column density are
given in \citet{tumlinson02}. The \ion{H}{1} column density
is derived from the 21 cm emission line and can be quite uncertain.
These authors estimated that the error on \ion{H}{1} was at least 33\%. 
When the upper-error on H$_2$ column density is large, $f_{\rm H_2}$ should
be considered a lower limit. Within the errors, $f_{\rm H_2}$ is 
similar to the Galactic value \citep{savage77} for similar total 
H$_2$ column density. 

Two main conclusions can be directly drawn from Figure~\ref{fig7}. 
(i) Not all the sight lines can be fitted with a single 
straight-line (Sk --65\degr22, AV 83, NGC346-3, Sk 188) and some 
are just at the edge of the error (NGC346-4, AV 327).
(ii) When levels $J=0,j$ ($j> 2$) can be fitted with a straight-line,
it is not generally the case that {\em all} the observed levels 
can be fitted with the same line (SK --67\degr69, BI 173, Sk --67\degr166, HD 269927, NGC346-6, Sk 82, AV 378),
i.e., at least two excitation temperatures are sometimes needed to describe properly
the distribution of the data points even when $N(J=0) < 10^{15}$ cm$^{-2}$. 
A single excitation temperature could reproduce the observed distribution
if either the $J=0$ level was less populated or the $J=1$ level more populated, which would
imply in both cases similar excitation temperatures to those observed in 
the Galaxy. This suggests non-equilibrium of the ortho to para ratio. 
Note that toward BI\,173 and HD\,269927
the $J=4$ level would still seem to be overpopulated to allow
a single excitation temperature to fit the observed distribution.

The future H$_2$ Galactic survey with \fuse\/ coupled with 
refined models and the Magellanic Clouds survey should improve our understanding
of H$_2$ in low density diffuse gas with different metallicities.

\newpage
\begin{table*}[!t]
\begin{center}
\caption{Atmospheric and other observational parameters of DI\,1388 and DGIK\,975}
\label{t0}
\begin{tabular}{lccccccccc}
\tableline\tableline
Star     &  $\alpha$ &  $\delta $ &  $V$   & $B-V$ & $T_{\rm eff}$ & $\log g$ & $v_{\rm t}$  & $v^\star_{\rm LSR}$ & $v\sin i$ \\
	& (J2000) & (J2000) & & & kK & dex & \km & \km & \km 
\\
\tableline
DI\,1388  	 &  02:57:11.94 &   $-$72:52:54.61  & 14.39 & $-0.26 $ & $32 $ &  $4.0 $  & $5 $ & $150 $ & $180 $ \\
DGIK\,975  	 &  04:19:58.63 &   $-$73:52:25.80  & 15.05 & $-0.12 $ & $20 $ &  $3.6$  &  $5 $  & $210 $ & $80 $ \\
\tableline
\end{tabular}
\tablecomments{Results are from \citet{hambly} (DI\,1388) and \citet{rol99} (DGIK\,975) and references therein.}
\end{center}
\end{table*}

\begin{deluxetable}{lccccccc}
\tablecolumns{7}
\tablewidth{0pc} 
\tablecaption{Atomic and molecular absorption lines in \fuse\/ spectra toward DI\,1388  \label{t1}} 
\tablehead{\colhead{Species}    &   \colhead{$\lambda_{\rm lab}$$^a$} &   \colhead{$f$$^a$} &
\colhead{S/N$^b$} &\colhead{$W_\lambda$} & \colhead{$\log N$} &\colhead{Note}\\ 
\colhead{}  &\colhead{(\AA)} & \colhead{} &\colhead{} &\colhead{(m\AA)}& \colhead{(${\rm cm}\, ^{-2}$)}&\colhead{}}
\startdata
\ion{C}{3}  	 & 977.020   &   $7.62 \times 10\, ^{-1}$	&     24	& $ 182.0\pm 15.1$	& $> 13.76$	  &  	       \\
\ion{N}{1} 	 & 953.655   &   $2.50 \times 10\, ^{-2}$	&     9		& $ 20.2 \pm 9.2$       & $14.07 \pm \, ^{0.17}_{0.26} $ &   	     \\
		 & 953.970   &   $3.48 \times 10\, ^{-2}$	&     10	& $16.8 :$		& $13.90 :$  &  		     	      \\
	         & 954.104   &   $6.76 \times 10\, ^{-3}$	&     9		& $ < 18 $	        & $<14.52  $		&  	     	      \\
	         & 964.626   &   $9.43 \times 10\, ^{-3}$	&     9		& $ 8.5:$	        & $14.05 : $		  &  	     	      \\
	         & 1134.415  &   $2.97 \times 10\, ^{-2}$	&     38	& $ 26.5 \pm 2.8$       & $> 13.97  $	  &  	 	     	    \\
	         & 1134.980  &   $4.35 \times 10\, ^{-2}$	&     18	& $ 33.3 \pm 4.9$       & $ > 13.93  $	  &  	 	     	    \\
\ion{N}{2}  	 & 1083.994  &   $1.15 \times 10\, ^{-1}$	&     12	& $ 118.9 \pm 9.3$      & $> 14.19  $	  &  	 	     	    \\
\ion{N}{3}  	 & 989.799   &   $1.23 \times 10^{-1}   $	&     7		& $ 136 \pm 22   $      & $ 13.5 :  $	  &  	 1	     	    \\
\ion{O}{1}  	 & 924.950   &   $1.54 \times 10\, ^{-3}$	&     10	& $ < 22 $		& $< 15.28 $	 	  &  	     		      \\
		 & 929.517   &   $2.29 \times 10\, ^{-3}$	&     13	& $ 21.8 \pm 3.9 $	& $15.31 \pm 0.17 $	  &  	      2\\
		 & 936.630   &   $3.65 \times 10\, ^{-3}$	&     19	& $ 34.9 \pm 4.4 $	& $15.18 \pm 0.06 $	  &   	     	    \\
		 & 948.686   &   $6.31 \times 10\, ^{-3}$	&     9		& $ 43   \pm 20  $	& $> 15.04: 	 $	  &   	     	     \\
		 & 950.885   &   $1.58 \times 10\, ^{-3}$	&     16	& $ 24.4 \pm 7.0 $      & $15.36 :$		  &          3 	    \\
	         & 1039.230  &   $9.20 \times 10\, ^{-3}$	&     20	& $ 60.4 \pm 5.4$       & $ >14.96  $		  &  	     	     \\
\ion{P}{2}  	 & 1152.818  &   $2.45 \times 10\, ^{-1}$	&     19        & $ 5.7: $	      	& $12.34 : $ &			     	    \\
\ion{S}{3}  	 & 1012.495  &   $4.43 \times 10\, ^{-2}$	&     22        & $ 14.3 \pm 4.0$	& $13.58 \pm \, ^{0.11}_{0.15}$ &			     	    \\
\ion{Ar}{1}  	 & 1048.220  &   $2.63 \times 10\, ^{-1}$	&     30        & $ 5.6 \pm 1.4$        & $12.39 \pm \, ^{0.09}_{0.13} $ &   		 \\
	  	 & 1066.660  &   $6.65 \times 10\, ^{-2}$	&     23        & $ < 9 $ 	      	& $ < 13.1 $		&     	     	     \\
\ion{Fe}{2}  	 & 1055.262  &   $7.50 \times 10\, ^{-3}$	&     24        & $ 7.9 \pm  3.9 $      & $14.05 \pm \, ^{0.17}_{0.27} $ &   		    \\
		 & 1063.176  &   $5.47 \times 10\, ^{-2}$	&     32        & $ 40.1\pm  3.8 $      & $13.92 \pm 0.05 $	&  	     		\\
		 & 1096.877  &   $3.20 \times 10\, ^{-2}$	&     22        & $ 25.8 \pm 5.5 $      & $13.93 \pm \, ^{0.09}_{0.12} $ &   		    \\
		 & 1125.448  &   $1.60 \times 10\, ^{-2}$	&     32        & $ 15.8 \pm 7.8 $      & $13.97 \pm \, ^{0.17}_{0.30} $ &   	    \\
		 & 1143.226  &   $1.77 \times 10\, ^{-2}$	&     32        & $ 14.1 \pm 3.3 $      & $13.87 \pm 0.10 $	&	     	    \\
		 & 1144.938  &   $1.06 \times 10\, ^{-1}$	&     24        & $ 73.0 \pm 4.6 $      & $13.90 \pm 0.03 $	&	     		    \\
\ion{Fe}{3}  	 & 1122.524  &   $5.44 \times 10\, ^{-2}$	&     24        & $ 24.5 \pm 3.7 $      & $13.66 \pm 0.07 $ &    	     	    \\
		 &  	     &					&	         & 		     &    				     	    \\
\hspace*{0.2cm} H$_2$ $J=0$  &  	     &   			     &	         & 		     &    		        		\\
1--0 $R(0)$	 & 1092.195  &   $5.90 \times 10\, ^{-3}$    &	17       & $20.6 \pm 5.2$    &  $14.88\pm 0.30$	    		        		 \\
3--0 $R(0)$  	 & 1062.882  &   $1.79 \times 10\, ^{-2}$    &	16       & $32.3 \pm 6.9$    &  \nodata	 	    		        		  \\
\hspace*{0.2cm} H$_2$ $J=1$  &  	     &   			     &	         & 		     &	    		        		 \\
0--0 $Q(1)$$^a$	 & 1009.771  &   $2.38 \times 10\, ^{-2}$    &	15       & $31.5 \pm 3.6$    &  $15.05 \pm 0.07$   		        		 \\
0--0 $R(1)$	 & 1108.633  &   $1.08 \times 10\, ^{-3}$    &	15       & $10.1 \pm 4.4$    &  \nodata	    			        	 \\
1--0 $R(1)$	 & 1092.732  &   $3.78 \times 10\, ^{-3}$    &	19       & $19.6 \pm 4.4$    &  \nodata	    			        	 \\
2--0 $R(1)$	 & 1077.697  &   $7.84 \times 10\, ^{-3}$    &	16       & $25.5 \pm 4.4$    &  \nodata	    			        	 \\
3--0 $R(1)$	 & 1063.460  &   $1.19 \times 10\, ^{-2}$    &	16       & $25.8 \pm 2.2$    &  \nodata	  	    		        		    \\
4--0 $R(1)$	 & 1049.960  &   $1.56 \times 10\, ^{-2}$    &	16       & $32.4 \pm 4.8$    &  \nodata	    			        	 \\
7--0 $R(1)$	 & 1013.435  &   $2.05 \times 10\, ^{-2}$    &	16       & $28.4 \pm 4.1$    &  \nodata	    			        	 \\
8--0 $R(1)$	 & 1002.449  &   $1.82 \times 10\, ^{-2}$    &	22       & $32.0 \pm 4.3$    &  \nodata	   	    		        		   \\
1--0 $P(1)$  	 & 1094.052  &   $1.97 \times 10\, ^{-3}$    &	19       & $14.3 \pm 4.9$    &  \nodata	 	    		        		  \\
2--0 $P(1)$  	 & 1078.923  &   $3.90 \times 10\, ^{-3}$    &	17       & $26.9 \pm 7.2$    &  \nodata	 	    		        		    \\
3--0 $P(1)$  	 & 1064.606  &   $5.66 \times 10\, ^{-3}$    &	12       & $22.3 \pm 7.3$    &  \nodata	 	    		        		   \\
4--0 $P(1)$  	 & 1051.031  &   $7.73 \times 10\, ^{-3}$    &	26       & $27.1 \pm 3.8$    &  \nodata	 	    		        		   \\
\hspace*{0.2cm} H$_2$ $J=2$  &  	     &   			     &	         & 		     &  		        				   \\
0--0 $Q(2)$	 & 1010.938  &   $2.45 \times 10\, ^{-2}$    &	18       & $21.0 \pm 3.5$    &  $14.60 \pm \, ^{0.22}_{0.16} $	        					   \\
1--0 $R(2)$  	 & 1094.244  &   $3.31 \times 10\, ^{-3}$    &	15       & $6.5 \pm  4.6$    &  \nodata	  	    		        		  \\
4--0 $R(2)$  	 & 1051.498  &   $1.40 \times 10\, ^{-2}$    &	23       & $23.0 \pm 3.5$    &  \nodata	  	    		        		  \\
8--0 $R(2)$  	 & 1003.984  &   $1.67 \times 10\, ^{-2}$    &	13       & $21.3 \pm 4.4$    &  \nodata	  	    		        		  \\
10--0 $R(2)$  	 & 983.589   &   $1.16 \times 10\, ^{-2}$    &	12       & $19.8 \pm 5.2$    &  \nodata	   	    		        		  \\
1--0 $R(2)$$^a$	 & 986.241   &   $2.62 \times 10\, ^{-2}$    &	11       & $26.5 \pm 5.2$    &  \nodata	    	    		        		    \\
\hspace*{0.2cm} H$_2$ $J=3$  &  	     &   			     &	         & 		     &  	    	        				  \\
0--0 $P(3)$  	 & 1014.504  &   $8.28 \times 10\, ^{-3}$    &	16       & $17.8 \pm 4.6$    &  $14.58 \pm \, ^{0.13}_{0.08} $	        					   \\
4--0 $P(3)$  	 & 1056.472  &   $9.56 \times 10\, ^{-3}$    &	25       & $18.0 \pm 3.0$    &  \nodata	   	    		        		  \\
1--0 $R(3)$$^a$	 & 987.445   &   $2.57 \times 10\, ^{-2}$    &	12       & $24.7 \pm 4.6$    &  \nodata	 	    		        		   \\
\hspace*{0.2cm} H$_2$ $J=4$  &  	     &   			     &	         & 		     &  		        				   \\
3--0 $R(4)$  	 & 1070.899  &   $9.67 \times 10\, ^{-3}$    &	20       & $<11$    &  $< 14.0 $	  	    		        		   \\

\enddata
\tablecomments{Uncertainties are $1\sigma$ errors. Upper limits indicate that no feature is present and
are $3 \sigma$ estimates. 
Lower limits indicate that the absorption line is saturated.
Colons indicate that the value is uncertain. For atomic species, the  column density is the direct 
integration of the apparent column density; while for H$_2$, it results from a COG analysis. Note however that the adopted
column densities for \ion{N}{1}, \ion{O}{1}, and \ion{Fe}{2} result from a COG analysis, see Table~\ref{t3} and \S~\ref{analysis}.
\\
($a$) Rest frame vacuum wavelengths and oscillator strengths are from Morton (private communication, 2000) for the atomic
and ionic species, except for the $f$-values of \ion{Fe}{2} which are from \citet{howk00}. 
The H$_2$ wavelengths are from \citet{abgrall93a,abgrall93b}, while
the H$_2$ $f$-values were calculated from the emission probabilities given by those
authors.
($b$) Signal-to-noise level per spectral resolution.
\\
(1) \ion{N}{3} $\lambda$989.8 is blended with \ion{Si}{2} $\lambda$989.9. The equivalent width is the
total strength of this feature (\ion{N}{3} and \ion{Si}{2}). To obtain the column density, the 
contribution of \ion{Si}{2} to the feature was removed (see \S~\ref{analysis} for more details).
(2) The red wing of the line is blended with local \ion{O}{1}. 
(3) The line is blended with local H$_2$.
(4) Blended with lower velocity components, profile integrated from 160 to 255 \km.
}
\end{deluxetable}

\begin{deluxetable}{lcccccc}
\tablecolumns{6}
\tablewidth{0pc} 
\tablecaption{Interstellar atomic/ionic absorption lines in \fuse\/ spectra toward DGIK\,975  \label{t2}} 
\tablehead{\colhead{Ions}    &   \colhead{$\lambda_{\rm lab}$$^a$} &   \colhead{$f$$^a$} &
\colhead{S/N$^b$} &\colhead{$W_\lambda$} & \colhead{$\log N_a$} \\ 
\colhead{}  &\colhead{\AA} & \colhead{} &\colhead{} &\colhead{(m\AA)}& \colhead{(${\rm cm}\, ^{-2}$)}}
\startdata
\ion{N}{1} 	 & 1134.165  &   $1.52 \times 10\, ^{-2}$    &	7       & $ < 30$       	& $< 14.2 $	  &  	 \\
	         & 1134.980  &   $4.35 \times 10\, ^{-2}$    &	9       & $ 31.9 \pm 9.7$       & $13.87 \pm \, ^{0.12}_{0.17} $	  &  	 \\
\ion{N}{2}  	 & 1083.994  &   $1.15 \times 10\, ^{-1}$    &	2       & \nodata		& \nodata	  &  	 \\
\ion{O}{1}  	 & 1039.230  &   $9.20 \times 10\, ^{-3}$    &	12      & $ 66.4 \pm 9.1$       & $ > 15.04  $		  &  	 \\
\ion{Si}{2}  	 & 1020.699  &   $1.64 \times 10\, ^{-2}$    &	6	& $ 34 : $		& $14.4 :$		  &  	 \\
\ion{P}{2}  	 & 1152.818  &   $2.45 \times 10\, ^{-1}$    &	8	& $< 26$		& $<13.0 $ &  	  \\
\ion{Ar}{1}  	 & 1048.220  &   $2.63 \times 10\, ^{-1}$    &	10	& $ <21 $		& $< 12.9$ &  	 \\
\ion{Fe}{2}  	 & 1096.877  &   $3.20 \times 10\, ^{-2}$    &	9	& $ 47.9 \pm 13.7 $	& $14.21 \pm \, ^{0.11}_{0.15} $ &  	  \\
		 & 1125.448  &   $1.60 \times 10\, ^{-2}$    &	12	& $ 23.6 \pm 6.2 $	& $14.20 \pm \, ^{0.10}_{0.13} $ &  	  \\
		 & 1143.226  &   $1.77 \times 10\, ^{-2}$    &	7	& $ 43.3 : $		& $14.39 : $	  &  	  \\
		 & 1144.938  &   $1.06 \times 10\, ^{-1}$    &	7	& $ 140 \pm 18 $	& $14.22 \pm 0.10 $	  &  	  \\
\ion{Fe}{3}  	 & 1122.524  &   $5.44 \times 10\, ^{-2}$    &	7	& $ 90.4 \pm 39.0 $	& $14.27 \pm \, ^{0.16}_{0.25} $ &  	 \\
\enddata
\tablecomments{Uncertainties are $1\sigma$ errors. Upper limits indicate that no feature is present and
are $3 \sigma$ estimates. 
Lower limits indicate that the absorption line is saturated.
Colons indicate that the value is uncertain.\\
($a$) Rest frame vacuum wavelengths and oscillator strengths are from Morton (private communication, 2000), 
except for the $f$-values of \ion{Fe}{2} which are from \citet{howk00}. 
($b$) Signal-to-noise level per spectral resolution.
}
\end{deluxetable}

\begin{deluxetable}{lccccccc}
\tablecolumns{8}
\tablewidth{0pc} 
\tablecaption{Highly ionized species toward DI\,1388  \label{t1a}} 
\tablehead{\colhead{Species}    &   \colhead{$\lambda_{\rm lab}$$^a$} &   \colhead{$f$$^a$} &
\colhead{$\log N$} &\colhead{$\log N$} &\colhead{$\log N$} &\colhead{$\log N$} &\colhead{Fit$^b$}\\ 
\colhead{}  &\colhead{(\AA)} & \colhead{} & \colhead{(${\rm cm}\, ^{-2}$)}&\colhead{(${\rm cm}\, ^{-2}$)}&\colhead{(${\rm cm}\, ^{-2}$)}&\colhead{(${\rm cm}\, ^{-2}$)}&\colhead{}\\ 
\colhead{}  &\colhead{} & \colhead{} & \colhead{MB}&\colhead{MB}&\colhead{HVC}&\colhead{MW}&\colhead{} \\
\colhead{}  &\colhead{} & \colhead{} & \colhead{$[155,245]$$^c$}&\colhead{$[100,245]$$^c$}&\colhead{$[70,100]$$^c$}&\colhead{$[-40,40]$$^c$}&\colhead{}}
\startdata
\ion{O}{6}  	 &  1031.926 	 &   $1.32 \times 10\,^{-1}$    & $ 13.13 \pm 0.16$   & $ 13.59 \pm 0.10$   & $ 13.18 \pm 0.12$  		& $ 14.08 \pm 0.05$	  	    &	P   \\
\ion{C}{4} 	 &  1548.195 	 &   $1.90 \times 10\,^{-1}$	& $ 13.27 \pm 0.06$   & $ 13.36 \pm 0.10$   & \nodata		 		& \nodata		  	    &	P   \\
	 	 &  1550.770 	 &   $9.52 \times 10\,^{-2}$	& $ 13.25 \pm 0.07$   & $ 13.28 \pm 0.10$   & $ 12.33 \pm\,^{0.15}_{0.23}$  	& $ 13.88 \pm 0.03$	  	    &	G   \\
	 	 &	    	 &   			    	& $ 13.34 \pm 0.13$   & $ 13.40 \pm 0.15$   & $ 12.45 \pm\,^{0.23}_{0.50}$	& $ 13.89 \pm 0.03$	  	    &	P  \\
	 	 &	    	 &   			    	& $ 13.33 :$	      & $ 13.54 :	$   & $ 12.98 :			 $	& $ 13.97 \pm 0.10$	  	    &	P  \\
\ion{Si}{4}      &  1393.755 	 &   $5.14 \times 10\,^{-1}$	& $ 12.87 \pm 0.06$   &  \nodata	    & $ 11.91 \pm 0.13$  		& $ 13.21 \pm 0.04$	  	    &	G   \\
		 &  		 &   			    	& $ 12.92 \pm 0.06$   & $ 13.02 \pm 0.08$   & $ 12.18 \pm 0.15$  		& $ 13.20 \pm 0.04$	  	    &	P   \\
		 &  		 &   			    	& $ 12.99 \pm 0.06$   & $ 13.18 \pm 0.08$   & $ 12.48 \pm 0.10$  		& $ 13.27 \pm 0.04$	  	    &	P   \\
 		 &  1402.770 	 &   $2.55 \times 10\,^{-1}$	& $ 12.92 \pm 0.06$   & \nodata		    & \nodata		  		& $ 13.29 \pm 0.04$	  	    &	G   \\
		 &  		 &   			    	& $ 12.90 \pm 0.05$   & $ 12.97 \pm 0.15$   & $ 12.20 \pm 0.20$			& $ 13.29 \pm 0.04$	  	    &	P   \\
		 &  		 &   			    	& $ 13.17 \pm 0.06$   & $ 13.40 \pm 0.07$   & $ 12.75 \pm 0.11$  		& $ 13.34 \pm 0.06$	  	    &	P   \\
\enddata
\tablecomments{Uncertainties are $1\sigma$ errors. Colons indicate that the value is uncertain. 
($a$) Rest frame vacuum wavelengths and oscillator strengths are from Morton (private communication, 2000).
($b$) Gaussian (G) or polynomial (P) fit to the continuum (see \S~\ref{hot} and Figure~\ref{fig5a0} for more details).
($c$) Range in \km\ over which the column density profile is integrated.}
\end{deluxetable}

\begin{deluxetable}{lccccc}
\tablecolumns{5}
\tablewidth{0pc} 
\tablecaption{Adopted column densities and ratios of the highly ionized species toward DI\,1388  \label{t1b}} 
\tablehead{\colhead{}  & \colhead{MB}&\colhead{MB}&\colhead{HVC}&\colhead{MW} \\
\colhead{}  & \colhead{$[155,245]$}&\colhead{$[100,245]$}&\colhead{$[70,100]$}&\colhead{$[-40,40]$ }}
\startdata
$N($\ion{O}{6})				& $ 13.13 \pm 0.16$  	 	& $ 13.59 \pm 0.10$		& $ 13.18 \pm 0.12$		& $ 14.08 \pm 0.05$ 	       \\
$N($\ion{C}{4})	    			& $ 13.29 \pm\,^{0.16}_{0.07}$  & $ 13.35 \pm\,^{0.20}_{0.10}$	& $ 12.39 \pm\,^{0.25}_{0.50}$	& $ 13.89 \pm 0.03$		       \\
$N($\ion{Si}{4})       			& $ 12.93 \pm\,^{0.20}_{0.07}$	& $ 13.18 \pm\,^{0.20}_{0.10}$	& $ 12.40 \pm\,^{0.35}_{0.30}$	& $ 13.27 \pm 0.07$ 	       \\
 $N($\ion{C}{4}$)/N($\ion{O}{6})	& $ 1.5 \pm\,^{1.5}_{0.6}$	& $ 0.6 \pm\,^{0.5}_{0.2}$	& $ 0.2 \pm\,^{0.2}_{0.2}$	& $ 0.6 \pm\,^{0.2}_{0.1}$ 	        \\
 $N($\ion{C}{4}$)/N($\ion{Si}{4})	& $ 2.3 \pm\,^{1.6}_{1.1}$	& $ 1.5 \pm\,^{1.5}_{0.8}$	& $ 1.0 \pm\,^{2.5}_{0.9}$	& $ 4.2 \pm\,^{1.2}_{0.8}$       \\
\enddata
\tablecomments{Error bars reflect the range of possible values. Numbers between 
brackets indicate the range in \km\ over which the column density profile is integrated.}
\end{deluxetable}

\clearpage

\begin{deluxetable}{lcc}
\tablecolumns{3}
\tablewidth{0pc} 
\tablecaption{Highly ionized species toward DGIK\,975  \label{t2a}} 
\tablehead{\colhead{}&\colhead{MB} &\colhead{MW}}
\startdata
$N($\ion{O}{6})	($\lambda$1031.926)	&   $ 14.26 \pm 0.12$		& $ 13.70 \pm\,^{0.17}_{0.27}$   \\
$N($\ion{C}{4})	($\lambda$1550.770)	&   $ 13.69 \pm 0.14$		& $ 13.90 \pm 0.10$   		 \\
$N($\ion{Si}{4}) ($\lambda$1393.755)	&   $ 13.42 \pm 0.06$		& $ 13.21 \pm 0.07$   		\\
$N($\ion{Si}{4}) ($\lambda$1402.770)	&   $ 13.40 \pm 0.13$		& $ 13.19 \pm 0.13$   		\\
$N($\ion{Si}{4}) (mean)			&   $ 13.41 \pm 0.10$		& $ 13.20 \pm 0.10$   		\\
$N($\ion{C}{4}$)/N($\ion{O}{6})		&   $ 0.3 \pm 0.1$		& $ 1.6 \pm 0.8$   		   \\
$N($\ion{C}{4}$)/N($\ion{Si}{4})	&   $ 1.9 \pm 0.8$		& $ 5.0 \pm 1.7$   		   \\
\enddata
\tablecomments{Uncertainties are $1\sigma$ errors.}
\end{deluxetable}

\begin{deluxetable}{lccccccccl}
\tablecolumns{10}
\tablewidth{0pc} 
\tablecaption{Summary of depletions of the MB toward  DI\,1388\label{t3}} 
\tablehead{ \colhead{Ions}    &   \colhead{}&   \colhead{} &   \colhead{MB}   &  \multicolumn{3}{c}{Galactic Depletions} 
& \colhead{SMC} & \colhead{LMC} & \colhead{Data} \\ 
\cline{5-7} \\ 
\colhead{$X^i$}  & \colhead{$\log\left(\frac{X}{{\rm S}}\right)_{\rm c}^a $} &   \colhead{$\log (N(X^i))$$^b$} & 
\colhead{$D(X)$}    & \colhead{Cold$^c$}   & \colhead{Warm$^c$}    & \colhead{Halo$^d$}& \colhead{$[X/{\rm S}]^e$} & \colhead{$[X/{\rm S}]^e$}& \colhead{}}
\startdata
\ion{C}{2}  			&  $+1.28$    &$> 14.87$ & $ > -0.65$  		        			& $-0.4$& $-0.4$& $(-0.4)$ 		& $-0.14$& $+0.06$			&  STIS  \\
\ion{N}{1}			&  $+0.70$    &$  14.20 \pm 0.11$ 	& $ -0.74 \pm \, ^{0.17}_{0.19} $ 		        & $-0.1$& $-0.1$& $(-0.1)$	& $-0.66$& $-0.26$   		& {\em FUSE} $+$  STIS \\
\ion{O}{1}			&  $+1.60$    &$  15.15 \pm 0.06$ 	& $-0.69 \pm \, ^{0.14}_{0.15} $				& $-0.4$& $-0.4$& $(-0.4)$	& $-0.16$& $+0.05$   	& {\em FUSE} $+$  STIS   \\
\ion{Mg}{2}			&  $+0.31$    &$< 15.00$ & $< +0.45$     					& $-1.2$& $-0.6$& $-0.3$ & 		  $+0.08$& $+0.44$		     	&  STIS  \\
\ion{Al}{2}			&  $-0.79$    &$> 12.82$ & $> -0.63$     					& $-2.4$& $-1.1$& $(-0.6)$ 		& $+0.20$& $+0.19$		     	& STIS   \\
\ion{Si}{2}			&  $+0.28$    &$  14.19 \pm 0.10$ & $-0.33 \pm \, ^{0.16}_{0.19}$    		& $-1.3$& $-0.4$& $-0.3$ 	& $+0.16$& $+0.02$			     	& STIS   \\
\ion{P}{2}			&  $-1.70$    &$12.34: $		& $-0.20:$				& $-0.5$& $-0.2$& $(-0.1)$	& \nodata& \nodata   				& {\em FUSE}   \\
\ion{Mn}{2}			&  $-1.74$    &$< 12.78$ & $< +0.28$    					& $-1.5$& $-1.0$& $-0.7$ 		& $+0.18$& $+0.25$		     	&  STIS\\
\ion{Ar}{1}			& $-0.71$&$12.39 \pm \, ^{0.09}_{0.13}$ & $-1.14 \pm \, ^{0.19}_{0.32}$				& (0.0)& (0.0)& (0.0)	& \nodata& \nodata   		&  {\em FUSE}  \\
\ion{Fe}{2}			&  $+0.24$    &$  13.94 \pm 0.02$ 	& $-0.54 \pm 0.14 $   			& $-2.2$& $-1.4$& $-0.6$ 	& $+0.01$& $+0.29$   				&{\em FUSE} $+$  STIS    \\
\ion{Ni}{2}			&  $-1.02$    &$  12.52 \pm 0.20$ & $-0.70 \pm \, ^{0.22}_{0.25} $   		& $-2.2$& $-1.4$& $-0.6$ 	& $+0.28$& $+0.36$			     	&  STIS  \\
\enddata
\tablecomments{$(a)$ Solar system meteoritic abundances from \citet{anders} except for C, N and O, which are
photospheric values from \citet{grevesse}; $\log({\rm S/H})_{\rm c} = -4.73$.
$(b)$ Adopted \ion{S}{2} column density, $\log(N({\rm S^+}))= 14.24 \pm 0.13$ dex. The errors on \ion{Si}{2} and \ion{S}{2}
were overestimated by $1/\sqrt2$ in \citet{lehner01a}.
$(c)$ Updated from \citet{jenkins87}, see \citet{lauroesch,welty97,welty99b}. 
$(d)$ From \citet{savage96,fitzpatrick}; and references
therein. Values in parentheses are estimated, see \citet{welty97,welty99b}.
For \ion{Ni}{2}, the depletions were corrected to take into account the new oscillator strength scaling, see \citet{lehner01a}.
$(e)$ $[X/{\rm S}]= \log(X/{\rm S})_{\rm SMC/LMC} - \log(X/{\rm S})_c$ (\citet{russell} but adjusted for C, N and O
photospheric values, and Al abundances are from \citet{welty97,welty99}, while LMC Si abundance is from \citet{korn},
but see also discussion in \citet{garnett}).}
\end{deluxetable} 

\clearpage

\begin{thebibliography}{}

\bibitem[Abgrall et al.(1993a)]{abgrall93a}
	Abgrall, H., Roueff, E., Launay, F., Roncin, J. Y., \& Subtil, J. L. 1993a, A\&AS, 101, 273

\bibitem[Abgrall et al.(1993b)]{abgrall93b}
	Abgrall, H., Roueff, E., Launay, F., Roncin, J. Y., \& Subtil, J. L. 1993b, A\&AS, 101, 323

\bibitem[Anders \& Grevesse(1989)]{anders}
	Anders, E., \& Grevesse, N., 1989, Geochim. Cosmochim. Acta, 53, 197

\bibitem[Barnes \& Hernquist(1992)]{barnes92}
	Barnes, J. E., \& Hernquist, L. 1992, \araa, 30, 705

\bibitem[Borkowski, Balbus, \& Fristrom(1990)]{borkowski90}
	Borkowski, K. J., Balbus, S. A.,  \& Fristrom, C. C. 1990, \apj, 355, 501


\bibitem[Christodoulou, Tohline, \& Keenan(1997)]{christ97}
	Christodoulou, D. M., Tohline, J. E.,  \& Keenan, F. P. 1997, \apj, 468, 810

\bibitem[Demers \& Battinelli(1998)]{demers98}
	Demers, S., \& Battinelli, P. 1998 \aj, 115, 154

\bibitem[Evans(1999)]{evans99} 
	Evans, N. J. II 1999, \araa, 37, 311

\bibitem[Fitzpatrick(1996)]{fitzpatrick} 
	Fitzpatrick, E. L. 1996, \apj, 473, L55

\bibitem[Grevesse \& Noels(1993)]{grevesse} 
	Grevesse, N., \& Noels, A. 1993, in Origin of the Elements, ed.
	N. Prantzos et al. (Cambridge: Cambridge Univ. Press), 15

\bibitem[Gardiner \& Noguchi(1996)]{gardiner}  
	Gardiner, L. T., \& Noguchi, M. 1996, \mnras, 278, 191

\bibitem[Garnett(1999)]{garnett} 
	Garnett, D. R. 1999, in IAU no. 190 New Views of the Magellanic Clouds,
	ed. Y.-H. Chu et al., p.266

\bibitem[Gibson et al.(2000)]{gibson00}
	Gibson, B. K., Giroux, M. L., Penton, S. V., Putman, M. E.,
	Stocke, J. T., \& Shull, J. M. 2000, \aj, 120, 1830

\bibitem[Hambly et al.(1994)]{hambly} 
	Hambly, N. C., Dufton, P. L., Keenan, F. P.,  
	Rolleston, W. R. J., Howarth, I. D., Irwin, M. J. 1994, \aap, 285, 716

\bibitem[Henry \& Worthey(1999)]{henry99} 
	Henry, R. B. C., \& Worthey, G. 1999  \pasp, 111, 919

\bibitem[Henry, Edmunds, \& K\"oppen(2000)]{henry00} 
	Henry, R. B. C., Edmunds, M. G., \& K\"oppen, J. 2000  \apj, 163, 165

\bibitem[Hollenbach, Werner, \& Salpeter(1971)]{hollenbach71} 
	Hollenbach, D. J., Werner, M. W., \& Salpeter, E. E. 1971  \apj, 163, 165

\bibitem[Hoopes et al.(2002)]{hoopes02} 
	Hoopes, C. G., Sembach, K. R., Howk, J. C., Savage, B. D., \& Fullerton, A. W. 2002, \apj, 569, 233

\bibitem[Howk et al.(2000)]{howk00} 
	Howk, J. C., Sembach, K. R., Roth, K. C., \& Kruk, J. W. 2000, \apj, 544, 867

\bibitem[Howk et al.(2002b)]{howk02a} 
	Howk, J. C., Savage, B. D., Sembach, K. R., \& Hoopes, C. G. 2002b, \apj, in press

\bibitem[Howk et al.(2002a)]{howk02} 
	Howk, J. C., Sembach, K. R., Savage, B. D., Massa, D., Friedman, S. D., \& Fullerton, A. W. 2002a, \apj, 569, 214

\bibitem[Irwin, Demers, \& Kunkel(1990)]{irwin90} 
	Irwin, M. J., Demers, S., Kunkel, W. E. 1990, \aj, 99, 191

\bibitem[Jenkins(1987)]{jenkins87} 
	Jenkins, E. B. 1987, in Interstellar Processes, edd. D.J. Hollenbach
	\& H.A. Thronson (Dordrecht: Reidel), 533

\bibitem[{Jenkins} {et al.}(2000a)]{jenkins00}
         Jenkins, E. B., et al. 2000a, \apj, 538, L81

\bibitem[{Jenkins} {et al.}(2000b)]{jenkins00a}
         Jenkins, E. B., Wo\`zniak, P. R., Sofia, U. J., Sonneborn, G., \& Tripp, T. M 
	 2000b, \apj, 538, 275

\bibitem[Kobulnicky \& Dickey(1999)]{kob99} 
	Kobulnicky, H. A., \& Dickey, J. M. 1999, \aj, 117, 908

\bibitem[Korn and Wolf(1999)]{korn} 
	Korn, A. J., \& Wolf, B. 1999, in IAU no. 190 New Views of the Magellanic Clouds, ed. Y.-H. Chu et al., p.241

\bibitem[Lauroesch et al.(1996)]{lauroesch} 
	Lauroesch, J. T.,  Truran, J. W., Welty, D. E., \& York, D. G. 1996, PASP, 108, 641 

\bibitem[Lehner(2000)]{lehner00} 
	Lehner, N. 2000, PhD Thesis, The Queen's University of Belfast

\bibitem[Lehner et al.(2001c)]{lehner01c} 
	Lehner, N.,  Fullerton, A. W., Sembach, K. R., Massa, D., Jenkins, E. B. 2001c, \apj, 556, L103 

\bibitem[Lehner, Keenan, \& Sembach(2001)]{lehner01b} 
	Lehner, N.,  Keenan, F.P., \& Sembach, K. R. 2001, \mnras, 323, 904 

\bibitem[Lehner et al.(2001)]{lehner01a} 
	Lehner, N.,  Sembach, K. R., Dufton, P. L., Rolleston, W. J. R., \& Keenan, F. P. 2001, \apj, 551, 781 

\bibitem[Lehner et al.(1999)]{lehner99} 
	Lehner, N., Sembach, K. R., Lambert, D. L., Ryans, R. S. I., \& Keenan, F. P. 1999, \aap, 352, 257

\bibitem[Lu, Sargent, \& Barlow(1998)]{lu98b} 
	Lu, L., Sargent, W. L. W., \& Barlow, T. A. 1998b, \aj, 115, 55

\bibitem[Lu et al.(1998)]{lu98} 
	Lu, L., Savage, B. D., Sembach, K. R., Wakker, B. P., Sargent, W. L. W., \& Oosterloo, T. A. 
	1998, \aj, 115, 162

\bibitem[{Mallouris} {et al.}(2001)]{mallouris01}
         Mallouris, C., et al. 2001 \apj, 558, 133

\bibitem[{Meyer} {et al.}(1998)]{meyer98}
         Meyer, D. M., Jura, M., \& Cardelli, J. A. 1998, \apj, 493, 222
	 
\bibitem[{Moos} {et al.}(2000)]{moos00}
         Moos, H. W., et al. 2000, \apj, 538, L1

\bibitem[Morton(1991)]{morton91} 
	Morton, D. C. 1991, \apjs, 77, 119

\bibitem[Prochaska et al.(2002)]{prochaska02} 
	Prochaska, J. X., Henry, R. B. C., O'Meara, J. M., Tytler, D., Wolfe, A. M.,
	Kirkman, D., Lubin, D., \& Suzuki, N., 2002 \apj, submitted

\bibitem[Putman(2000)]{putman00} 
	Putman, M. E. 2000, PASA, 17, 1

\bibitem[Putman et al.(2002)]{putman02} 
	Putman, M. E., Staveley-Smith, L., Freeman, K. C., Gibson, B. K., \& Barnes, D. G. 
	2002, \apj, to be submitted
 
\bibitem[Richter et al.(2001)]{richter01}
	 Richter, P., Sembach, K.R., Wakker, B.P., \& Savage, B.D. 2001, \apj, 526, L181 

\bibitem[Rolleston et al.(1999)]{rol99}
	 Rolleston, W. R. J., Dufton, P. L., McErlean, N. D., \& Venn, K. A. 1999, \aap, 348, 728
	 
\bibitem[Russell and Dopita(1992)]{russell} 
	Russell, S. C., \& Dopita, M. A. 1992, \apj, 384, 508

\bibitem[{Sahnow} {et al.} (2000)]{sahnow00}
        Sahnow, D. J., et al. 2000, \apj, 538, L7

\bibitem[Savage et al.(1977)]{savage77} 
	Savage, B. D., Bohlin, R. C., Drake J. F., \& Budich, W. 1977, \apj, 216, 291

\bibitem[Savage \& Sembach(1991)]{savage} 
	Savage, B. D., \& Sembach, K. R. 1991, \apj, 379, 245
	
\bibitem[Savage \& Sembach(1996)]{savage96} 
	Savage, B. D., \& Sembach, K. R. 1996, \araa, 34, 279
	
\bibitem[Savage et al.(2000)]{savage00} 
	Savage, B. D. et al., 2000, \apj, 538, L30

\bibitem[Scoville, Sanders, \& Clemens(1986)]{scoville86} 
	Scoville, N. Z., Sanders, D. B., \& Clemens, D. P. 1986, \apj, 310, L77

\bibitem[Sembach et al.(2000)]{sembach00} 
	Sembach, K. R., Howk, J. C., Ryans, R. S. I., \& Keenan, F. P. 2000, \apj, 528, 310

\bibitem[Sembach et al.(2001)]{sembach01} 
	 Sembach, K .R., Howk, J. C., Savage, B. D., \& Shull J. M. 2001, \apj, 121, 992 

\bibitem[Sembach \& Savage(1992)]{sembach92} 
	Sembach, K. R., \& Savage, B. D. 1992, \apjs, 83, 147

\bibitem[Sembach, Savage, \& Hurwitz(1999)]{sembach99} 
	Sembach, K. R., \& Savage, B. D., Hurwitz, M. 1999, \apj, 524, 98

\bibitem[Shull \& Beckwith(1982)]{shull82}
         Shull, J. M., \& Beckwith, S. 1979, \apj, 227, 131

\bibitem[Shull \& McKee(1979)]{shull79}
         Shull, J. M., \& McKee, C.F. 1979, \apj, 227, 131

\bibitem[{Shull} {et al.}(2000)]{shull00}
         Shull, J. M., et al. 2000, \apj, 538, L73

\bibitem[Slavin, Shull, \& Begelman(1993)]{slavin93}
         Slavin, J. D., Shull, J. M., \& Begelman, M. C. 1993, \apj, 407, 83

\bibitem[Smoker et al.(2000)]{smoker00} 
	 Smoker, J.V., Keenan, F.P., Polatidis, A., Mooney, C.J.,  Lehner, N., \&
		Rolleston, W.R.J. 2000, \aap, 363, 451 

\bibitem[Sofia \& Jenkins(1998)]{sofia98} 
	Sofia, U. J., \& Jenkins, E. B. 1998, \apj, 499, 951  

\bibitem[Spitzer(1996)]{spitzer96} 
	 Spitzer, L. 1996, \apj, 458, L29

\bibitem[Spitzer, Cochran, \& Hirshfeld(1974)]{spitzer74} 
	 Spitzer, L., Cochran, W. D., \& Hirshfeld, A. 1974, \apjs, 28, 373

\bibitem[Sutherland \& Dopita(1993)]{sutherland93} 
	 Sutherland, R. S., \& Dopita, M. A. 1993, \apjs, 88, 253

\bibitem[Tripp \& Savage(2000)]{tripp00} 
	 Tripp, T. M., \& Savage, B. D. 2000, \apj, 542, 42

\bibitem[Tumlinson et al.(2002)]{tumlinson02} 
	 Tumlinson, J., et al. 2002, \apj, 566, 857

\bibitem[Vila-Costa \& Edmunds(1993)]{vila93} 
	 Vila-Costa, M. B., \& Edmunds, M. G. 1993, \mnras, 265, 199

\bibitem[Wakker \& van Woerden(1997)]{wakker97} 
	Wakker. B. P., \& van Woerden, H., 1997, ARA\&A, 35, 217

\bibitem[Welty et al.(1997)]{welty97} 
	Welty, D. E., Lauroesch, J. T., Blades, J. C.,  Hobbs, L. M., \& York, D. G. 1997, \apj, 489, 672  

\bibitem[Welty et al.(1999a)]{welty99} 
	Welty, D. E., Frisch, P. C., Sonneborn, G., \&  York, D. G. 1999a, \apj, 512, 636  

\bibitem[Welty et al.(1999b)]{welty99b} 
	Welty, D. E., Hobbs, L. M., Lauroesch, J. T., Morton, D. C., Spitzer, L., \& York, D. G. 1999b, \apjs, 124, 465  

\bibitem[Wiese, Bonvallet, \& Lawler(2002)]{wiese02} 
	Wiese, L. M., Bonvallet, G. A., \& Lawler, J. E.  2002, \apj, 569, 1032  

\bibitem[Zsarg\'o et al.(2002)]{zsargo02} 
	 Zsarg\'o, J., Sembach, K. R., Howk, J. C., \& Savage, B. D. 2002, \apj, in preparation

\end{thebibliography}
\end{document}